\documentclass[1p,times]{elsarticle}

\usepackage{ecrc}

\newcommand{\nchunks}{ {B} }
\newcommand{\ncustomers}{ {n} }
\newcommand{\bncustomers}{ {\bf n} }
\newcommand{\bNCUSTOMERS}{ {\bf N} }
\newcommand{\punsust}{ {p} }

\newcommand{\realav} { {v} }

\newcommand{\available} { {V} }
\newcommand{\unavailable} { {V} }
\newcommand{\calunavailable} { {\mathcal{V}} }

\newcommand{\sustained}{available among the peers}
\newcommand{\unsustained}{unavailable among the peers}
\newcommand{\unsustainedl}{unavailable among the leechers}

\newcommand{\bsustained}{blocks available among the peers}
\newcommand{\tunavailable}{unavailable}
\newcommand{\tavailable}{available}

\newcommand{\among}{among the peers}

\newcommand{\bN}{ {\bf N} }

\newcommand{\bS}{ {\bf S} }

\newcommand{\bOmega}{ {\mbox{\boldmath $\Omega$}} }

\newcommand{\btau}{ {\mbox{\boldmath $\tau$}} }

\newcommand{\bvarep}{ {\mbox{\boldmath $\varepsilon$}} }

\newcommand{\bzr}{ {\bf 0} }
\newcommand{\bon}{ {\bf 1} }

\newcommand{\be}{ {\bf e} }

\newcommand{\bmm}{ {\bf m} }

\newcommand{\bs}{ {\bf s} }

\input{epsf}

\usepackage{amsmath}
\usepackage{amssymb}
\usepackage{amsfonts}
\usepackage{comment}
\usepackage{bm}
\usepackage{comment}
\usepackage{url}
\usepackage{mathrsfs}
\usepackage{latexsym}
\usepackage{verbatim}
\usepackage{url}
\usepackage{mathrsfs}
\usepackage{amsmath}

\newtheorem{theorem}{Theorem}[section]

\newproof{pf}{Proof}

\newtheorem{lemma}{Lemma}[section]

\newtheorem{definition}{Definition}[section]
\newtheorem{example}{Example}[section]
\newlength{\labelexample}
\volume{00}

\firstpage{1}

\journalname{Performance Evaluation}

\runauth{}

\jid{PEVA}

\jnltitlelogo{Performance Evaluation}

\usepackage{amssymb}

\usepackage[figuresright]{rotating}

\usepackage{algorithm}
\usepackage{algorithmic}

\newcommand{\eat}[1]{}

\begin{document}

\begin{frontmatter}



\dochead{}

\title{Estimating Self-Sustainability  in Peer-to-Peer Swarming Systems}


\author[label1]{Daniel S. Menasch\'{e}}
\ead{sadoc@cs.umass.edu}
\author[label2]{Antonio A. A. Rocha}
\ead{arocha@land.ufrj.br}
\author[label2]{Edmundo A. de Souza e Silva}
\ead{edmundo@land.ufrj.br}
\author[label2]{Rosa M. Le\~{a}o}
\ead{rosam@land.ufrj.br}
\author[label1]{Don Towsley}
\ead{towsley@cs.umass.edu}
\author[label1]{Arun Venkataramani}
\ead{arun@cs.umass.edu}
\address[label1]{Computer Science Department, University of Massachusetts, Amherst, MA, United States}
\address[label2]{COPPE/PESC, Federal University of Rio de Janeiro, Rio de Janeiro, RJ, Brazil}

\begin{abstract}
Peer-to-peer swarming is one of the \emph{de facto} solutions for distributed content dissemination in today's Internet.  By leveraging resources provided by clients, swarming systems reduce the load on and costs to publishers.  However, there is a limit to how much cost savings can be gained from swarming; for example, for  unpopular content peers will always depend on the publisher in order to complete their downloads.  In this paper, we investigate such a dependence of peers on a publisher.  For this purpose, we propose a new metric, namely \emph{swarm self-sustainability}.   A swarm is referred to as self-sustaining if all its blocks are collectively held by peers; the self-sustainability  of a swarm is the fraction of time in which the swarm is self-sustaining.    We pose the following question: how does the self-sustainability of a swarm vary as a function of  content popularity, the service capacity of the users, and the size of the file? We present a model to answer the posed question.  We then propose efficient solution methods to compute self-sustainability.  The accuracy of our estimates is validated against simulations. Finally, we also provide   closed-form expressions for  the fraction of time that a given number of blocks is  collectively held by peers.
\end{abstract}


\end{frontmatter}


\section{Introduction}

Peer-to-peer swarming, such as used by \eat{the immensely popular} BitTorrent~\cite{bt}, is a  scalable and efficient way to publish   content in today's Internet.  Peer-to-peer swarming has been widely studied during the last decade, and its use by enterprises is steadily growing~\cite{warner, twitter, amazon, ubuntu}.  \eat{In essence, the popularity of peer-to-peer swarming occurs because it allows  publishers to  provide a large number of files to clients, at a low cost. }  By leveraging resources provided by clients, peer-to-peer swarming  decreases costs to publishers, and provides scalability and system robustness.  As  demand for content increases, system capacity scales accordingly, as all clients collaborate with each other while downloading the desired content.  As the demand for multimedia files and the size of these files increase, peer-to-peer swarming systems have become an important content dissemination solution for many content providers~\cite{amazon, ubuntu}. 
 
 \eat{
Although peer-to-peer swarming systems are a powerful tool for content delivery, the  vast majority  of publishers  adopts client-server strategies.  Indeed, there are still a number of challenges to be addressed before peer-to-peer swarming systems will be widely adopted.   \eat{For instance, publishers have worried that the use of peer-to-peer systems can hinder simple billing and copyright enforcement mechanisms.    Native support for electronic sales and advertisement in peer-to-peer systems~\cite{vuze} is an initial step to meet the monetization needs, whereas automated auditors~\cite{sherman, p2ppolice} and digital right mechanisms (DRM)~\cite{drm} have been suggested to meet the copyright requirements.   

}  }

However, there is a limit on how much savings can be gained from swarming techniques.  For example, in the case of unpopular content, peers must  rely on the publisher in order to complete their downloads. In this paper, we investigate such a dependence of peers on a publisher. \eat{, willing to  provide an extensive catalog of titles, that include unpopular ones~\cite{anderson}.  }

A swarm is a set of peers interested in the same content (file or bundle of files~\cite{mainarticle}) that exchange \eat{(download and upload) } blocks of the files among themselves.    We consider a scenario where each swarm includes one stable publisher that is always online and ready to serve content.  The corresponding system is henceforth referred to as a hybrid peer-to-peer system, since peers can always rely on the publisher if they cannot find blocks of the files  among themselves. \eat{ (content is always available). } If all blocks are \sustained, \eat{excluding the publisher, } the swarm is referred to as \emph{self-sustaining}.   Quantifying \eat{the } swarm self-sustainability, defined as the fraction of time during  which the swarm is self-sustaining, is useful for provisioning purposes.  The larger the swarm's self-sustainability, \eat{which depends on the popularities of the files and the capacity of the peers,} the lower the dependency of peers on the publisher, and the lower the  bandwidth needed by the publisher to serve~the~peers.  \eat{Given a bandwidth capacity constraint at  the publisher,  swarm self-sustainability can also be used for dimensioning purposes, to determine the number of swarms that the publisher is able to serve. \eat{handle.  }}

The primary contribution of this paper is  a model to study   \eat{the dependency of peers on  publishers in peer-to-peer swarming systems.  The dependecy is measured as the } swarm  self-sustainability.    We use a two-layer model to quantify  swarm \eat{fraction of time in which a swarm is self-sustaining } self-sustainability as a function of the number of blocks in the file, the mean upload capacity of peers and the popularity of a file.   The upper layer of our model  captures how  user dynamics evolve over time, while  the lower layer  captures the probability of a given number of blocks being \sustained{}   conditioned on a fixed upper layer population state.  Our model is flexible enough to account for large or small numbers of blocks in the file, heterogeneous download times for different blocks, and peers residing in the system after completing their downloads.   We derive closed-form expressions for the distribution of the number of \bsustained{}   \eat{ to solve the proposed model.  }     \eat{Then, we provide probabilistic interpretation for the closed-form expressions,}  and apply them to show that \eat{for a given mean user capacity and file popularity,  the} self-sustainability increases as a function of the number of blocks in the file. \eat{ Nevertheless, the closed-form } The derived expressions involve sums and subtractions of large numbers, and are amenable to numerical errors.   Hence,  we present an efficient algorithm to compute the swarm self-sustainability that  avoids these problems.
We \eat{apply our solution algorithm to } then numerically investigate the minimum popularity needed to attain a given self-sustainability level. \eat{, and observe that the log of the minimum  popularity is linear in the log of the file size.   } Finally, we validate the estimates made by the model against detailed simulations.

The remainder of this paper is organized as follows.  After providing a brief background into swarming systems in~\textsection\ref{sec:primer},  \eat{ and exemplify using measurements some of the difficulties that may be faced by publishers relying on swarming systems to disseminate content. } \eat{in~\textsection\ref{sec:motivate} we further motivate the problem considered in this paper  and} in \textsection\ref{sec:model} we propose our model.  In \textsection\ref{sec:instant} \eat{and \textsection\ref{sec:bawsteady} } we present an efficient algorithm to solve the proposed model followed by analytical results in~\textsection{\ref{sec:analytical}}. \eat{, respectively. } In \textsection\ref{sec:exp} we evaluate our model against experiments. In \textsection\ref{assumptions} we discuss some limitations and caveats of our model,   \textsection\ref{sec:related} presents the related work and \textsection\ref{sec:conclusion} concludes the paper.

\section{Swarming Systems Primer}

\label{sec:primer}

A swarm is a set of peers concurrently sharing content of common interest.  Content might be a file or a bundle of files that are distributed together.   The content is divided into  blocks that peers upload to and download from each other.   Since there is no interaction between peers across swarms,  each swarm can be \eat{analyzed individually.  } studied separately.

  \eat{download the file at rate $\mu$ chunks/second and } \eat{therefore users are indifferent to the identities of the  chunks they have.  \eat{If a user  has $h$ 
chunks of a file, it does not matter which of the ${\nchunks \choose h}$  combination of chunks he may have. \eat{, where ${\nchunks \choose h} = {\nchunks!}/[{h! (\nchunks-h)!}]$. } }}

BitTorrent is one of the most popular applications that uses peer-to-peer swarming for content dissemination, and we will use it to  illustrate how swarming works.  Unlike a traditional server-based system, BitTorrent includes  a {\em tracker} that  promotes the interaction of participating peers. The identities of the trackers are announced to peers in {\em torrent } files, which can be found and downloaded through search engines such as Torrent Finder~\cite{torrentfinder}.   \eat{When a peer joins the system, the tracker serves a {\em torrent} file that contains information about chunks constituting the file. } Peers periodically query the tracker to obtain a random subset of other peers  in the swarm in order to exchange (upload and download) blocks with them.   Peers also discover new neighbors from other peers, in addition to the tracker, when the Peer Exchange (PEX) extension is enabled.

There are two kinds of peers in the system: {\em seeds} and {\em leechers}. Seeds are peers that have completed their  downloads and only upload blocks. Leechers are peers that have not completed their downloads and are actively  downloading (and uploading) blocks of the file. Thus, leechers turn into seeds upon completing their downloads. Leechers adopt a tit-for-tat incentive strategy while downloading the file, i.e., leechers preferentially upload content to  other leechers that reciprocate likewise, and ``choke'' or ignore leechers that do not reciprocate. 

As many leechers  may leave the system immediately after completing the download, content publishers often  support a stable seed  that we refer to as the {\em publisher}. \eat{The availability of BitTorrent depends on the availability of } A publisher is guaranteed to have all of the blocks constituting the file.   In the rest of this paper we assume that each swarm includes   one publisher. \eat{ per swarm. }\eat{, which we refer to as the (stable) publisher.} \eat{and use the terms {\em seed} and {\em publisher} interchangeably to refer to that seed, as well as {\em peers} and {\em users} to refer to the other members of the~swarm. }
\eat{
\subsection{Measuring Chunk Distribution}

since we have the bitmaps, we can do a measurement inspired by~\cite{piece, piecetr}
}

\eat{\paragraph{Block Selection Strategy}}

\eat{In } BitTorrent peers adopt the rarest first policy to decide which blocks to download from their neighbors.  According to the rarest first policy,  a peer prioritizes the rarest blocks when selecting the ones to download next. \eat{ from its neighbors.  } We say that a peer is \emph{interested} in another peer if the latter can provide blocks to the former.  Since  rarest first guarantees a high diversity of blocks in the system,  any peer is almost always interested in any other peer, which in general  yields high system performance~\cite{legoutrf}.    

Note, however, that in BitTorrent peers only have local information about the system.  Hence, they can only implement a  \emph{local} rarest first policy. \eat{ can be deployed.} \eat{ in practice. }  The intrinsic limit on the number of connections that a user can establish naturally provides each of them with only a myopic view of the system.    Firewalls, NATs and other exogenous factors may also prevent users from establishing connections among themselves.  \eat{ Under these circumstances, users that uniformly and independently select their chunks correspond to the extreme case in which  no peer can communicate with another, and rarest first corresponds to the other extreme of the spectrum, under which all users can communicate with each other.  \emph{Local} rarest first, as implemented in BitTorrent~\cite{legoutrf}, lies somewhere in between. }

\eat{\input{why}}

\section{Model}

\label{sec:model}

In this section, we present our model to estimate  swarm self-sustainability. \eat{ to disseminate a file in a swarm. } Our model is hierarchical.  The upper layer characterizes  
user dynamics, and the lower layer comprises a performance model used to quantify the distribution of \bsustained, for a given  state of the upper layer model.\footnote{Each layer of the model is self-containted.  In Appendix~\ref{app:integr} we provide an alternate description of our model, integrating the user dynamics into the lower layer and explicitly capturing the evolution of the blocks owned by each user (users signatures) in time.}  	We  present each of the layers, in \textsection\ref{sec:uplayer} and \textsection\ref{sec:lowlayer}, respectively, \eat{discuss the assumptions in \textsection\ref{assumptions} } and then introduce the metric of interest in \textsection\ref{sec:metrics}.

\subsection{User Dynamics Model}

\label{sec:uplayer}

\begin{figure} 
\center
\includegraphics[scale=0.5]{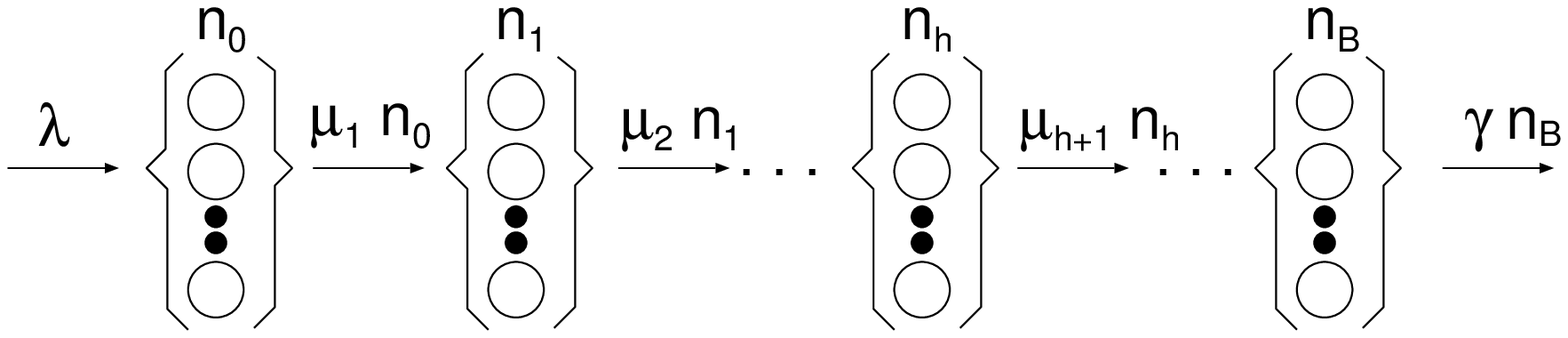} 
\caption{User dynamics. In stage $h$, there are $\ncustomers_h$ users, each user owning $h$ blocks, $0 \leq h \leq \nchunks$. } \label{tandem}
\end{figure}

\eat{\textbf{Assumption 1: } }
 \eat{
Our workload consists of peers that arrive according to a Poisson process with rate $\lambda$, the file popularity.   }

\medskip
A file consists of $\nchunks$~blocks. Requests for a file arrive according to a Poisson process with rate~$\lambda$. \eat{,  the file popularity.}  We further assume that the time required for a user to download its $j^{th}$ block is a  random variable with mean $1/\mu_{j}$, $1 \leq j \leq B$.  After completing their downloads, peers remain in the system for a time with mean  $1/\gamma$.

We model user dynamics with  \eat{a tandem Jackson network of} $(\nchunks+1)$~M/G/$\infty$ queues in series.  \eat{, where each M/G/$\infty$ queue models the download of a block.     } Each of the first $B$~M/G/$\infty$ queues models the download of a block, and   capture the self-scaling
property of BitTorrent swarms, i.e., each peer
brings one unit of service capacity to the system.  The last queue  captures the residence time of seeds (see Figure~\ref{tandem}).

The system state is characterized by a
$(\nchunks+1)$-tuple, $\bncustomers=(\ncustomers_0, \ncustomers_1, \ncustomers_2, \ldots, \ncustomers_{\nchunks})$,   where $\ncustomers_h$ represents the number of customers in queue $h$,   i.e., the number of users that
have downloaded $h$ blocks of the file, $0 \leq h \leq \nchunks$.     We denote by $\bNCUSTOMERS$ the random variable characterizing the current state of the upper layer model  and by $\bncustomers$ its realization.    The number of peers  in the system is referred to as $n$, $n=\sum_{i=0}^{B} n_i$. \eat{ and $|\bncustomers|$.  }

Peers arrive according to a Poisson process with rate $\lambda$  to queue 0 and 
transit from queue $h$,  also referred to as stage $h$,  to queue $h+1$ ($0 \le h \le \nchunks-1$) with rate $\mu_{h+1}$, the download rate of the $(h+1)^{th}$ block downloaded by a peer.  The mean residence time in queue $\nchunks$ captures the mean time that peers remain in the system
after completing their downloads, $1/\gamma$. Setting $\gamma = \infty$
models the case where all peers immediately  leave the system after completing the
download.  Throughout this paper, unless otherwise stated,  we assume that \eat{$\gamma=\infty$ and that } the mean download times of all blocks are the same, $1/\mu_j=1/\mu$, $1 \leq j \leq B$.  Nevertheless, all results are easily extended to the case where \eat{$\gamma < \infty$ and } the mean time   it takes for a user to download its $j^{th}$ block is $1/\mu_j$.

Let 	$\pi(\ncustomers_0,\ldots,   n_{\nchunks})$ be the joint steady state population probability distribution, $\pi(n_0, \ldots, n_B)=P(\bN = (n_0, \ldots, n_B))$, of finding $\ncustomers_h$ users in the $h^{th}$ queue, $0 \leq h \leq \nchunks$, and let $\pi_h(\ncustomers_h)=P(N_h=n_h)$, $h=0,\ldots, B$,  be the corresponding marginal probability. The steady state distribution of the queueing system has the following  product form,
\begin{equation} \label{eqpi}
\pi(\ncustomers_0,\ldots,  \ncustomers_{\nchunks-1},  \ncustomers_{\nchunks}) = \prod_{h=0}^{\nchunks} \pi_h(\ncustomers_h) =   \frac{(\lambda/\gamma)^{\ncustomers_B}}{\ncustomers_B!} e^{-\lambda/\gamma}    \prod_{h=0}^{\nchunks-1}  \left(                             \frac{\rho^{\ncustomers_h}}{\ncustomers_h!} e^{-\rho}  \right)
\end{equation}
 where $\rho =  \lambda/\mu$ is  \eat{$\mu_i = \mu$ for $0 \le i < \nchunks$ and $\mu_{\nchunks} = \gamma$.  } the {\em load} of the system  (refer to Table~\ref{tab:notation} for notation).

\subsection{Performance Model For a Given Population State}

\label{sec:lowlayer}

{
\begin{table}[t!] 
\center 
\begin{tabular}{ll} 
\hline
\hline 
\multicolumn{2}{l}{parameters} \\
\hline
$\lambda$ & mean arrival rate of peers (peers/s) \\
$1/\mu$ & mean time to download a block (s) \\
$\rho=\lambda/\mu$ & mean load of the system (per stage) \\
\eat{$1/\ncustomers$ & mean residence time of seeds (s) \\}
$\nchunks$ & number of blocks in file  \\
$1/\gamma$ & mean residence time of seeds \\
\eat{$\eta$ & publisher bandwidth per chunk   \\
&delivered (KBps/chunk)\\}
\hline
\multicolumn{2}{l}{variables}  \\
\hline
\eat{ $h$ & upper layer stage, $0 \leq h \leq \nchunks$  \\ }
$\ncustomers_h$ & number of users that own $h$ blocks \\
${\bncustomers}{=} {(\ncustomers_0}{,} {\ldots}, {\ncustomers_{\nchunks})}$ & upper layer state  \\
$\pi(\bncustomers) \eat{{=}  \prod_{{h}{=}{0}}^{\nchunks }\pi_h(\ncustomers_h)}$ & steady state probability of state $\bncustomers$ \\
$n= \eat{|\bncustomers| =} \sum_{i=0}^{\nchunks} \ncustomers_i$  & number of peers in the system \\
	\eat{
\multicolumn{2}{l}{variables characterizing chunk distribution (lower layer) }\\
\multicolumn{2}{l}{ \quad when the upper layer state is $\bncustomers$ } 
 \\
\hline
$s_{\bncustomers, h,r}$ & bit vector of $r^{th}$ user in stage $h$ \\
$s_{\bncustomers}$ & lower layer state  \\
\eat{
$n_{\bncustomers,h} =  \ncustomers_h h$ & copies of chunks owned by users \\
& who  have $h$ chunks each \\
$n_{\bncustomers} = \sum_{h=0}^\nchunks n_{\bncustomers,h}$ & copies of chunks in the system \\ }
\eat{$C'_{\bncustomers}(r)$ &  same as above, in case users  may have \\ 
& multiple replicas of the same chunk \\ } 
\hline }
\eat{ $C(l)$ &  number of  blocks that have $l$ replicas among peers \\ }
\eat{ $A{=}P(\unavailable = 0)$ & chunk availability  \\ }
$\available$ & number of \bsustained  \\
\hline
\multicolumn{2}{l}{metrics} \\
\hline
\eat{$\unavailable_{\bncustomers} {=}  C_{\bncustomers}(0)$ & number of unavailable chunks \\ }
\eat{ $\calunavailable$ & $\#$ of unavailable chunks in steady state \\ }
\eat{ $\mathcal{A}$ \eat{=P(\calunavailable = 0)$} & chunk availability in steady state \\ }
\eat{$W_{\bncustomers} = \eta \unavailable_{\bncustomers}$  & instant. publisher bandwidth (KBps) \\
$\mathcal{W} = \eta \calunavailable$ & publisher bandw. in steady state (KBps) \\}
$p(\realav) = P(\unavailable=\realav)$ & probability of $\realav$ blocks being available among peers\\
${A}{=}p(\nchunks)$ & swarm self-sustainability  \\
\hline
\hline
\end{tabular} 
\caption{Table of notation. Vectors are denoted by bold face symbols.   \eat{When the context implies $\bncustomers$, we drop the subscript $\bncustomers$ from the variables.} \eat{Metrics conditioned on the upper layer state and on steady state are denoted by plain and stylized letters, respectively. } \eat{Metrics conditioned on the upper layer state and on steady state are denoted by plain and stylized letters, respectively.  } Unless otherwise stated, $\gamma=\infty$ in which case $n_B = 0$. \eat{ and, with a slight abuse of notation,  $\bncustomers \in \mathbb{N}^{B}$ rather than $\bncustomers \in \mathbb{N}^{B+1}$. } When referring to block availability, it is subsumed availability {\em among peers} (excluding publisher).   }
\label{tab:notation}
\end{table}
}

\label{sec:chunkdistrmodel}

\eat{We now describe the instantaneous performance model.      }

We now describe the  lower layer of the model.  
Given the current population state, $\bncustomers=(\ncustomers_0, \ldots, \ncustomers_{\nchunks})$, our goal  is to determine the distribution of the number of \bsustained.   \eat{.  To this aim, we consider the simple  \emph{uniform model}, } \eat{as described next.    }
We begin by stating our key modeling assumption.

\textbf{Uniform and independent block allocation: } In steady state, the set of blocks owned by a randomly selected user in stage $h$ is chosen  uniformly at random among the ${B \choose h}$ possibilities  and independently among users.

\eat{
Although we do not account for the exact dynamics of how the signature of each user  evolves over time, in order to describe our model it is convenient to $(i)$  consider  an arbitrary ordering of the users and $(ii)$ to associate to each user a bit vector  which characterizes the blocks the user has.    }
A user $u$ in stage $h$,  $0 \leq h \leq \nchunks$,  has  a signature $s_{h,u} \in \{ 0, 1 \}^\nchunks$,  defined as a $\nchunks$ bit vector where the $i^{{th}}$ bit is set to 1 if the
user has block $i$ and 0 otherwise.  Each user in stage $h$  owns $h$ blocks and has one of ${\nchunks \choose h}$ possible signatures.   \eat{
Denote by $\btau$ a vector whose $h^{th}$ element is  $h$. The \eat{it follows from~\eqref{en} that}   number of chunk replicas in state $\bncustomers$ is $\bncustomers \cdot \btau$.  }

Under the uniform and independent block allocation, signatures are chosen uniformly at random and independently among users; the latter is clearly a strong assumption since in any peer-to-peer swarming system the signatures of users are correlated.  Nevertheless,  in~\textsection\ref{sec:evalmain} we show that the effect of such correlations on our metric of interest, swarm self-sustainability, is negligible in many interesting scenarios.  Therefore, we proceed with our analysis under such an assumption.    \eat{even under such an independence assumption our model still accurately captured the behavior of simulated swarms.  }

   Let $S_{h,u}$ be the random variable denoting the signature of the $u^{th}$ user in stage $h$, and $s_{h,u}$ its realization, $1 \leq u \leq n$. 
The sample space of the lower layer model,  $\Omega_{\bncustomers}$, for a given state of the upper layer, $\bncustomers$, is the set of  all $\{ 0,1 \}^{B n }$ bit vectors in which element $B (u-1) + i$ equals one if the $u^{th}$ user has block $i$, and zero otherwise,  $1 \leq u \leq n$,  $1 \leq i \leq B$.  An element in $\Omega_{\bncustomers}$  is the concatenation of $n$ bit vectors of size $B$ each.   $\Omega_{\bncustomers}$ has cardinality 
$|\Omega_{\bncustomers}| = \prod_{h=0}^{\nchunks} {{\nchunks \choose h}}^{\ncustomers_h}
$. 
Then, under the uniform and independent block allocation assumption,  
\begin{equation} \label{uniformjoint}
P( S_{1,1} {=} s_{1,1}{,} \ldots{,} S_{B,\ncustomers_B} {=} s_{B,\ncustomers_B} | \bNCUSTOMERS = \bncustomers) \eat{{=} \prod_{h=0}^{B} \prod_{i=1}^{\ncustomers_h} P( S_{h,i} {=} s_{h,i} ) }  {=}  \frac{1}{|\Omega_{\bncustomers}|}
\end{equation}
In the next section, we relate the upper and lower layer models, showing how ~\eqref{eqpi} and ~\eqref{uniformjoint} yield the key metric of interest, namely, swarm self-sustainability.

\subsection{Self-Sustainability}

\label{sec:metrics}

We now define the key metric of interest, swarm self-sustainability.  Let $V$ denote the steady state number of \bsustained. \eat{, $\unavailable$. } \eat{  Content unavailability among peers is related to the utilization and the bandwidth consumption of the publisher since, in an hybrid peer-to-peer swarming system, peers contact the publisher if they can't get the content from each other. }
\eat{hen the  population state is ${\bncustomers}$. }  \eat{It is given as} \eat{Both $\unavailable$ and $\mathcal{A}$ depend on the scheduling policy adopted by the peers  to exchange chunks, and   are further discussed in the upcoming subsections. 
Once the chunk distribution policy is specified, } \eat{The self-sustainability of the swarm, in steady state,  is}
\eat{$\mathcal{A}$,}      
Denote by $\punsust(\realav)$ the steady state probability that $\realav$ blocks are~available~among~the~peers, \eat{$\punsust(\realav)=P(V=\realav)$, }  \eat{.   The steady state probability that $\realav$ blocks are \sustained{} is}
\begin{equation}   \label{naivev}
p(\realav) =  P(V=\realav)=\sum_{\bncustomers \in \mathbb{N}^{B+1} \eat{= (\ncustomers_0, \ncustomers_1, \ldots, \ncustomers_{\nchunks})}} P(V=\realav|\bNCUSTOMERS={\bncustomers}) \pi(\bncustomers)  
\end{equation}

\begin{definition} \label{def:sesu} The swarm self-sustainability, ${A}$, is  the steady-state    probability that the  peers have the entire file, 
\begin{equation}
A=p(B)
\end{equation}
\end{definition}

Definition~\ref{def:sesu} together with equation~\eqref{naivev} yield, \eat{

Conditioning $P(V=B)$ on the upper layer state, the swarm self-sustainability is }
\begin{equation}  \label{naivea}
{A} =  \sum_{\bncustomers \in \mathbb{N}^{B+1} \eat{= (\ncustomers_0, \ncustomers_1, \ldots, \ncustomers_{\nchunks})}} P(V=\nchunks|\bNCUSTOMERS={\bncustomers}) \pi(\bncustomers)  {=}  \sum_{\bncustomers \in \mathbb{N}^{B+1} \eat{= (\ncustomers_0, \ncustomers_1, \ldots, \ncustomers_{\nchunks-1})}} P(V=\nchunks|\bNCUSTOMERS={\bncustomers})  \frac{(\lambda/\gamma)^{\ncustomers_B}}{\ncustomers_B!} e^{-\lambda/\gamma}      \prod_{h=0}^{\nchunks-1} \left( \frac{\rho^{\ncustomers_h}}{\ncustomers_h!} e^{-\rho}  \right) 
\end{equation}
The second equality in ~\eqref{naivea} follows from {\eqref{eqpi}}.   $P(V=\nchunks|\bNCUSTOMERS=\bncustomers)$ is obtained from~\eqref{uniformjoint} (see Appendix~\ref{sec:apppv}).

\eat{\textbf{The Impact of Altruistic Lingering}}

\eat{Note that in the definition of self-sustainability we assume that $\gamma=\infty$.  }

If peers leave the system immediately after concluding their downloads, we refer to the swarm self-sustainability as $A_\infty$. \eat{$A=A(\infty)$, and we also drop the explicit dependence of $\punsust(\realav;\gamma)$ and  $P(V=\nchunks|\bNCUSTOMERS={\bncustomers};\gamma)$ on $\gamma$.}
The swarm self-sustainability,   $A$,   when $\gamma < \infty$, is obtained from   $A_{\infty}$  as follows,  
\begin{equation} \label{relaainf}
A = 1-(1-A_{\infty})\exp(-\lambda/\gamma), \qquad \gamma < \infty
\end{equation}
The above follows because a block is \tunavailable{} among the peers if it is \unsustainedl{} and there are no seeds in the system.    \eat{If $\gamma=\infty$, we drop the dependency of $A$ on $\gamma$, and $A=A(\infty)$ is given by \eqref{naivea}.   }

Note that $A$ is expressed through~\eqref{naivea}  as an infinite sum. \eat{the sum in the definition of self-sustainability has an infinite number of terms because the number of users in a swarm may be arbitrarily large.  } In what follows, we approximate $A$ \eat{the self-sustainability  } by its truncated version, $A^{(N)}$, considering only population states in which there are no more than  $N$ users in the system,      \eat{so that $\sum_{n=0}^{N} e^{-\rho B} (\rho B)^n/n! \ge 1- \eta$.  }
\begin{equation}  \label{naiveatrunc}
{A}^{(N)} =  \sum_{\bncustomers \in \mathbb{N}^{B} s.t. n \leq N} P(V=\nchunks|\bNCUSTOMERS={\bncustomers}) \pi(\bncustomers)  
\end{equation}
The value of $N$ is based on the desired error tolerance $\eta$, and is chosen as described at the end of~\textsection\ref{sec:bwratio}.  
A naive  use of~\eqref{naiveatrunc} yields an inefficient algorithm to compute~${A^{(N)}}$ \eat{and $\beta$ } by exploring a number of states that grows exponentially with respect to the file size. \eat{However,  }
This problem is addressed in the next section, where we provide an efficient algorithm to evaluate $A^{(N)}$.

In the rest of this paper, we refer  to the truncated  self-sustainability, $A^{(N)}$, simply as self-sustainability, the distinction between $A^{(N)}$ and $A$ \eat{is not } being clear from the context.    In addition, since $A$ \eat{, $\gamma < \infty$, } is readily obtained from $A_{\infty}$ using~\eqref{relaainf}, henceforth we focus on the case $\gamma=\infty$ and make the dependence of $p(v)$ on $\gamma$ explicit, denoting it by $p(v;\gamma)$,  whenever $\gamma < \infty$.

\section{An Efficient Solution Algorithm to Evaluate Self-Sustainability}
\label{sec:instant}
\label{sec:bwratio}

In this section we present an efficient algorithm to compute \eat{~\eqref{naiveatrunc} } swarm self-sustainability in polynomial time. \eat{, $O(NB)$.}  \eat{and leading to computational requirements that are combinatorial in the number of blocks in the file.  However, in \textsection\ref{sec:bwratio} we show a recursion to compute ${A}$  (and, in general, $P(V=v)$)  in polynomial time, for all the scenarios of interest. } The key insight consists of aggregating the states in the upper layer of the model in  such a way that the lower layer metrics are computed once per aggregate rather than once per state.     The algorithm relies on three observations about our model, the first related to the performance model for a given population state (lower layer model) and the last two related to the user dynamics (upper layer model).

Let $\psi_h(k,\realav)$ be the probability that, in a system in which $k$ blocks are initially available among peers,  $\realav$ blocks become \sustained{} after an additional user contributes $h$  blocks. \eat{ contributed by the additional user. }\eat{ contributes $h$ blocks.} \eat{a user when previously there were $i$ \sustained.  } Then, $\psi_h(k,\realav)$ is characterized by a hypergeometric distribution,
\begin{equation} \label{psiimh}
\psi_h(k,\realav) = \frac{{k \choose h-(\realav-k)}   {\nchunks-k \choose \realav-k} }{{\nchunks \choose h}}, \quad k=0, 1, \ldots, B-1,  B, \quad  \realav = \max(k,h), \ldots, B-1, B
\end{equation}
Equation~\eqref{psiimh}  follows because  there are ${\nchunks-k \choose \realav-k}$ ways in which $\realav-k$ blocks of the additional user  do not overlap with the $k$ previously \tavailable{} blocks,  and there are ${k \choose h-(\realav-k)} $  ways in which the other $h-{\realav+k}$ blocks can overlap  with previously \tavailable{}   blocks (see Figure~\ref{fig:recursion}).  
A recursion to compute $\psi_h(k,\realav)$ is presented in Appendix~\ref{sec:apprecpsi}. \eat{The  recursion is convenient to avoid numerical problems since it only involves sums and  products of probabilities, i.e., numbers between zero and one.}

\begin{figure}[t] 
\center
\includegraphics[scale=0.6]{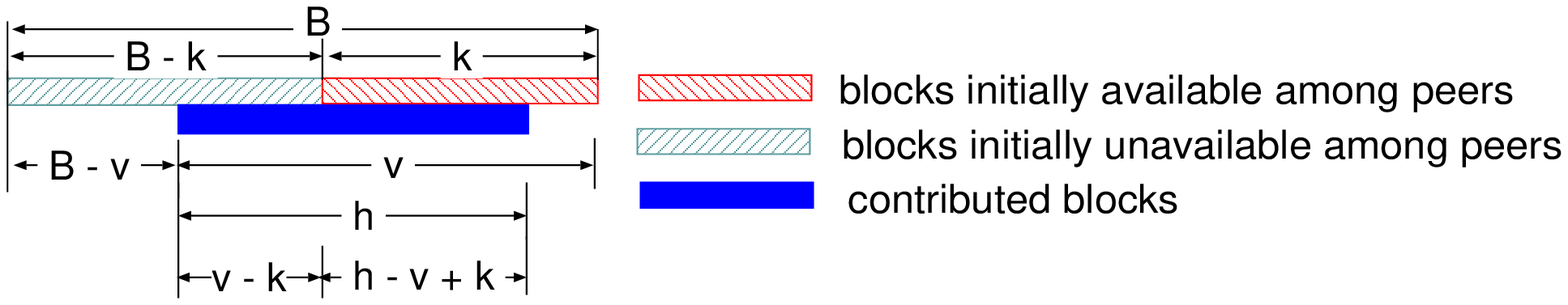}
\caption{Recursion to compute probability of $\realav$ blocks being \unsustained.  There are initially $k$ blocks \unsustained, and $\realav$ after  a user contributes $h$ blocks.} \label{fig:recursion}
\end{figure}

Our second observation regards the steady state probability that a randomly selected user is in stage $h$, denoted by ${\sigma}(h)$ ($0 \leq h \leq B-1$). It can be shown that  ${\sigma}(h)=1/B$, $0 \leq h \leq B-1$.  This is a consequence of the assumption that the download times of all blocks have the same mean, $1/\mu$. \eat{  (see~\textsection{\ref{assumptions}}). } Note that, in general,  if a user downloads its $(h+1)^{th}$ block at rate $\mu_{h+1}$, $0 \leq h \leq B-1$, the model can be easily parameterized by setting ${\sigma}(h)=(1/\mu_{h+1})/(\sum_{i=1}^{B} 1/\mu_i)$.

Our third and last observation concerns the total number of users in the system. The total population  is characterized by a Poisson random variable with mean $B \lambda/\mu$. This follows from the fact that the sum of $B$ independent Poisson random variables, with mean $\lambda/\mu$, is a Poisson random variable with mean $B \lambda/\mu$.

Denote by $\punsust_n(\realav)$
the probability that $\realav$ blocks are \sustained{} conditioned on the presence of $n$ users in the system,
\begin{equation} \label{umpvm}
\punsust_n(\realav) = P(V=\realav \Big| |\bncustomers|=n)
\end{equation}
It follows from the discussion in the previous paragraph that $\punsust(\realav) = \sum_{n=0}^{\infty} \punsust_n(\realav) e^{-B \rho} (B\rho)^n  / n!
$. 
 The truncated version of $p(\realav)$, $p^{(N)}(\realav)$, is
\begin{equation} 
\punsust^{(N)}(\realav) = \sum_{n=0}^{N} \punsust_n(\realav) e^{-B \rho} (B\rho)^n  / n!
\end{equation}
It remains to show how to compute $\punsust_n(\realav)$.  This is accomplished by making use of our first two observations, as summarized by the following lemma. \eat{using Finally, our first two observations yield the following recursion to compute $\punsust_h(\realav)$.}

 \begin{lemma} The probability of $\realav$ \tavailable{} blocks, conditioned on $n$ users in the system, $\punsust_{n}(\realav)$, satisfies the following recursion,  \label{lemmaprob}
 \begin{equation}
\punsust_n(\realav) {=} \left\{  
\begin{array}{ll}  
\displaystyle \sum_{h=0}^{\min(\realav,\nchunks-1)} \displaystyle  \sum_{k=\realav-h}^{\realav} \punsust_{n-1}(k) \psi_h(k,\realav) /{B}, & n \ge 1 \\
1, & n =0, \realav = 0 \\
0, & n =0, \realav \neq 0
\end{array} \right. \label{mainrec}
\end{equation}
\end{lemma}
In Appendix~\ref{sec:applemma}, we show that~\eqref{mainrec}  correctly computes~\eqref{umpvm}.  Next, we further simplify recursion~\eqref{mainrec}.  
Changing the order of the summations and adapting their limits accordingly \eat{(see Figure~\ref{region})} yields
\eat{
\begin{equation} \label{simplified1}
\punsust_n(\realav) {=} \left\{  
\begin{array}{cc}  
 \displaystyle  \sum_{k=m}^{B} \displaystyle \punsust_{n-1}(k) \sum_{h=k-\realav}^{\nchunks-1}  \psi_h(k,\realav) /{B} & n \ge 1 \\
1 & i =0, \realav = B \\
0 & i =0, \realav \neq B
\end{array} \right. 
\end{equation}}
\begin{equation} \label{simplified1}
\punsust_n(\realav) {=}  \sum_{k=0}^{\realav} \displaystyle \punsust_{n-1}(k) \sum_{h=v-k}^{\nchunks-1}  \psi_h(k,\realav) /{B},\qquad n \ge 1 
\end{equation}
The base cases are $\punsust_0(\realav) {=} 1$ if $\realav = 0$ and  $\punsust_0(\realav) {=} 0$  if $m \neq 0$.
\eat{
Note that the initial range of $(h,k)$ is $( 0 \ldots B-\realav, \realav \ldots \min(\realav+h,B))$. Therefore, $k \leq m+h$ which implies that $k -\realav \leq h$.  The range is equivalent to $(k-\realav \ldots B-\realav, \realav \ldots B)$ . }
\eat{
\begin{figure}
\center
\includegraphics{region}
\caption{Range of $(h,k)$ is given by $h=0 \ldots B-\realav$ and $k=\min(B,m+h)$ (bold line) or equivalently $k=m \ldots B$ and $h=k-\realav \ldots B-\realav$. }
\label{region}
\end{figure}
}
In Appendix~\ref{sec:appderiv1ok} we derive the following result, \eat{(for instance, using the Sigma~\cite{sigma} package -- see details below)} 
\begin{equation} \label{1okp1}
\sum_{h=m-k}^{\nchunks-1}  \psi_h(k,\realav)  = 
\left\{ 
\begin{array}{ll}
(B+1)/ (B-k+1), & 0 \le m \le B-1\\
k/(B-k+1), & \realav = B
\end{array} \right.
\end{equation}
Equation \eqref{1okp1} \eat{, which was obtained with the help of the  Sigma~\cite{sigma} package,} is key to further simplifying \eqref{simplified1}. Replacing \eqref{1okp1} into \eqref{simplified1} yields, after algebraic manipulation,

\eat{
\begin{equation}
\punsust_n(\realav) {=} \left\{  
\begin{array}{cc}  
 \displaystyle  \sum_{k=m}^{B} \displaystyle \punsust_{n-1}(k) (B+1)/(B(k+1))  & n \ge 1, m >1 \\
  \displaystyle  \sum_{k=0}^{B} \displaystyle \punsust_{n-1}(k) (B-k)/(B(k+1))  & n \ge 1, \realav=0 \\
1 & n =0, \realav = B \\
0 & n =0, \realav \neq B
\end{array} \right.
\end{equation}
which, after manipulation, can be shown to be equivalent to the following recursion which has complexity that grows linearly on $B$,
}

\begin{theorem} The probability of $\realav$ blocks being \sustained, $\punsust(\realav)$, equals  \label{theo:efficientalgo}
\begin{equation} \label{eq:sumnottru}
\punsust(\realav) = \sum_{n=0}^{\infty} \punsust_n(\realav) e^{-B \rho} (B \rho)^n  / n!
\end{equation}
where $\punsust_n(\realav)$ satisfies the following recursion, \eat{ correctly computes  \eat{0 \leq n \leq N,} } $ 0 \leq \realav < B$,
\begin{equation} \label{dimsimple}
\punsust_n(\realav) {=} \left\{  
\begin{array}{ll}  
1/B^n, & n \ge 1, \realav=0 \\
 \displaystyle  \punsust_n(\realav-1)+\punsust_{n-1}(\realav)((B+1)/(B(B-\realav+1))), & n \ge 1,  B > \realav  > 0 \eat{ \\
1-\sum_{l=0}^{B-1} \punsust_n(l), & n \ge 1,   \realav = \nchunks \\ } \\
1, & n =0, v=0 \\
0, & n =0, v\neq0 
\end{array} \right.
\end{equation}
\eat{The base cases are $p_0(\realav)=1$ if $\realav = 0$ and $p_0(\realav)=0$ if $\realav \neq 0$.  } \eat{The probability of all blocks being available among the peers when there are $n$ users in the system,  $\punsust_n(B)$, is}  \eat{probability $p_n(B)$ is }  and $\punsust_n(B)=1-\sum_{\realav=0}^{B-1} \punsust_n(\realav)$.  The approximation $p^{(N)}(v)$ for~\eqref{eq:sumnottru} is obtained by truncating  the infinite sum at $N$ and is computed in time $O(NB)$.    \eat{For this purpose,  use the following recursion to compute }\end{theorem}

\eat{
{\em 1) Lower Bound on Availability. }  The recursion proposed in Theorem~\ref{theo:general} has as its base case queue $\nchunks$.  Values are computed   moving across all queues, from right to left in Figure~\ref{tandem}.  
Now assume that our  goal is solely to obtain a lower bound on the probability of all chunks being available.  Since $a_{h}(0)$ increases as $h$ decreases,  one can proceed as follows.       Compute $a_{B-1}(0), a_{B-2}(0), \ldots, a_{1}(0)$, in that order, stopping when the desired availability level is reached.  If the  distribution of the number of users is the same across all queues, proceeding that way we consider first the queues that contribute most to the availability. 
}

\eat{{\em 2) Truncating the Sum. }  }

Theorem~\ref{theo:efficientalgo} yields an efficient algorithm to evaluate  \eat{truncated version of the } swarm self-sustainability.  \eat{$\punsust^{(N)}(\nchunks)=A^{(N)}$.}   \eat{Note that since the algorithm involves truncating the infinite sum~\eqref{eq:sumnottru} at integer $N$, it actually computes an approximation to $\punsust(\realav)$. }  
\eat{Recall that $N$ is the maximum number of users allowed in the system. }    The algorithm \eat{implied by Theorem~\ref{theo:efficientalgo}}  has complexity $O(NB)$, since $\punsust_n(\realav)$ is computed for $0 \leq n \leq N$ and $0 \leq m \leq B$.    Note also that once the elements $\punsust_n(\realav)$ are computed for a fixed $B$, one can obtain the self-sustainability for different values of $\rho$ in time $O(B)$.

\eat{Given that the expected number of users in the system is $\rho B$,  it is convenient to write the maximum allowed number of users $N$ as $N = \alpha \rho B$ ($\alpha >0$), where $\alpha$ is a proportionality constant. } Let $\bvarep(N)$ be the truncation error, ${\bvarep}(N)=p(B)-p^{(N)}(B)$.  
The maximum number of users in the system, $N$, can be chosen as a function of $\bvarep(N)$,
\begin{equation}
\bvarep(N)=\sum_{n=N+1}^{\infty}  \punsust(\realav) e^{-B \rho} (B\rho)^n  / n! \leq \sum_{n=N+1}^{\infty}   e^{-B \rho} (B\rho)^n  / n! = 1 -\sum_{n=0}^{N}   e^{-B \rho} (B\rho)^n  / n! 
\end{equation}
If $\rho B$ is large ($\rho B > 1000$), the Poisson distribution is well approximated by a normal distribution.  In this case, $N$ can be chosen so that $1-\Phi((N-B\rho)/\sqrt{B\rho}) \leq \eta$, where $\eta$ is the desired error tolerance and $\Phi(\cdot)$ is the standard normal cdf.

Theorem~\ref{theo:efficientalgo} assumes $\mu_h = \mu$, $0 \leq h \leq B-1$. 
If that is not the case,   self-sustainability can be computed in time $O(NB \log B)$ using an alternative recursion.   We refer the reader to~Appendix~\ref{sec:apphet} for details.

\section{Model Analysis}

\label{sec:analytical}

We derive closed-form expressions for the probability that $\realav$ blocks are \sustained{} in the system and for the mean number of \tavailable{} blocks,  in~\textsection\ref{closedform1} and ~\textsection\ref{closedform2},  respectively. \eat{, and provide probabilistic interpretations of the expressions.  }
\eat{
In this section we provide closed-form expressions to $\punsust_n(\realav)$. } The closed-form expressions are useful in order to gain insight on how different system parameters impact self-sustainability.  In~\textsection\ref{closedform2} we use the closed-form expressions to compute the minimum popularity to attain a given self-sustainability level.  In~\textsection\ref{closedform3} we show that self-sustainability increases with file size.    However, the closed-form expressions  may lead to numerical problems, if used to compute the self-sustainability for large files (e.g., $B>500$),  since they involve sums and subtractions of large numbers.  This is why the recursion presented in~\textsection\ref{sec:instant} is useful. \eat{above, in contrast,  is convenient to avoid numerical problems, since it requires only sums and multiplications of probabilities.  }

In order to simplify the closed-form expressions, in the remainder of this  section \eat{we remove the self-regarding hypothesis (see~\textsection\ref{assumptions}) and} we  assume that $\gamma=\mu$, i.e.,  peers,   after completing their  downloads, linger  in the system as seeds for an interval with duration drawn from an exponential distribution with mean $1/\gamma$.    Recall that when $\gamma$ is finite, we make the dependence of  $p_n(\realav)$ on $\gamma$ explicit, and denote it by $p_n(\realav;\gamma)$.     \eat{Henceforth, we focus on $p_n(\realav;\mu)$.   }

\subsection{Self-Sustainability Closed-Form Expression} \label{closedform1}

\eat{Therefore, we can compute the metrics of interest in linear time.  }

  \eat{

Assuming that $\gamma=\mu$, and} Similar arguments to those in ~\textsection\ref{sec:instant} (Theorem~\ref{theo:efficientalgo}) yield, for $\gamma = \mu$ and $0 \leq n \leq N, 0 \leq v \leq B$, 
\begin{equation} \label{dimsimple2} 
\punsust_n(\realav; \mu) {=} \left\{  
\begin{array}{ll}  
1/(B+1)^n, & n \ge 1, \realav=0 \\
 \displaystyle  \punsust_n(\realav-1;  \mu)+\punsust_{n-1}(\realav; \mu)(1/(B-\realav+1)), & n \ge 1,  0 < \realav  \leq B \\
1, & n=0, v=0 \\
0, & n=0, v\neq0 \\
\end{array} \right.
\end{equation}
\eat{The base cases are $\punsust_0(0; \mu)=1$ and $\punsust_0(\realav; \mu)=0$ ($\realav \neq 0$).  }
This recursion can be solved (see~Appendix~\ref{sec:appderclosed}), to obtain \eat{ we show that \eqref{dimsimple2} admits closed-form solution,}\footnote{The expression corresponding to equation~\eqref{closedform} for the case $\gamma=\infty$ is found in Appendix~\ref{sec:apppnv}.}  \eat{ref{sec:apppnv}.}
\eat{\begin{equation} \label{closedform}
\punsust_n(\realav) {=} \left\{  
\begin{array}{cc}  
(1/(B+1))^n & n \ge 1, \realav=B  \\
{B \choose \realav} \sum_{l=0}^{B-\realav} {B-\realav \choose l} (-1)^l (\realav+l+1)^{-n} & n \ge 1, 0 \leq \realav < B
\end{array} \right.
\end{equation} }
\begin{equation} \label{closedform}
\punsust_n(\realav; \mu) {=} 
{B \choose \realav} \sum_{l=0}^{\realav} {\realav \choose l} (-1)^l (B-\realav+l+1)^{-n}, \qquad 1 \le n, \qquad 0 \leq \realav \leq B
\end{equation} 
In particular, the probability that all blocks are \sustained, conditioned on the number of users in the system, is $
\punsust_n(\nchunks; \mu)= \sum_{l=0}^{B} {B \choose l} (-1)^l (l+1)^{-n} 
$.  Using this expression we derive, in Appendix~\ref{sec:d0} the corresponding unconditional probability, namely,  the swarm self-sustainability, \eat{ is derived in , and equals}
\begin{equation} \label{d0}
\punsust(\nchunks; \mu) = \sum_{l=0}^{B} {B \choose l} (-1)^l e^{-(B+1) \rho l / (l+1)} = 1-\sum_{l=1}^{B} {B \choose l} (-1)^{l+1} e^{-(B+1) \rho l / (l+1)}
\end{equation}
We now interpret~\eqref{d0} using the inclusion/exclusion principle, which allows us to apply the Bonferroni inequalities in~\textsection\ref{sec:minloadanl} and ~\textsection\ref{closedform3}.  \eat{It can be shown that} \eat{For this purpose, note} The term \eat{the rightmost summation in~\eqref{d0} corresponds to the probability that at least one block is \unsustained.  \eat{This probability, in turn, can be computed using the inclusion/exclusion principle.  } The term  } $\exp({-(B+1) \rho l / (l+1)})$ is  the probability that $l$ specific (\emph{tagged}) blocks are \unsustained{}  (refer to~Appendix~\ref{taggedblocks} for  the derivation).  So, ${B \choose l} \exp({-(B+1) \rho l / (l+1)})$ is the sum of the probabilities  that \emph{any  } $l$ blocks are \unsustained.  Therefore, as a consequence of the inclusion/exclusion principle, the probability that \emph{at least} one block is \unsustained{} equals the rightmost summation in ~\eqref{d0}, and $\punsust(\nchunks; \mu)$ is its complement.

In what follows, we use the above closed-form expression to analyze the mean number of blocks \unsustained{} as well as the impact of the file size on the self-sustainability.

\subsection{Minimum Load to Attain Self-Sustainability} \label{closedform2}

\label{sec:minloadanl}

\eat{In this section} We now provide a simple expression to estimate the minimum load necessary to attain high self-sustainability.  The result relies on approximating  swarm  self-sustainability using the mean number of \tavailable{} blocks \among. \eat{, which is derived next.  }
 The mean number of \tavailable{} blocks \among, $E[V]$, is   
 $E[V] = \nchunks(1 -  q)$, where $q$ is the probability that a tagged block is \unsustained, 
\begin{equation} \label{esteadyv}
q =\exp({-\rho(\nchunks +1)/2})
\end{equation}
The expression of $q$ is readily obtained from ~\eqref{d0}.  Note  that the mean number of unavailable blocks, $Bq$,  equals the first term $(l=1)$ in the rightmost summation in ~\eqref{d0}.  This observation coupled with an application of the Bonferroni inequality~\cite{bonferroni} to ~\eqref{d0} implies that $1-p(B;\mu) \leq B-E[V]$.    When $E[V]\approx B$, the upper bound $B-E[V]$  provides an approximation to the fraction of time that the swarm is not self-sustaining, $1-p(B;\mu)$.   \eat{This approximation is  used in the end of this section to compute the load necessary to attain a given self-sustainability level.   }

\eat{
The derivation yields, in addition to $q$, the probability that a tagged block is unavailable among peers when $\gamma=\infty$, denoted as $q_{\infty}$.  
}

Next, we  present a simple  alternative derivation of~\eqref{esteadyv}. 
  \eat{ For this purpose, further insight into~\eqref{esteadyv}  is obtained by}   Let $q_{\infty}$ be the  probability that a tagged block is \tunavailable{} among leechers (excluding seeds).  In order to compute $q_{\infty}$, note that the mean time that a leecher holds a tagged block is $\sum_{l=0}^{B-1} l/(\mu B) = (B-1)/(2 \mu)$ and the rate at which leechers acquire a tagged block is given by a Poisson process with mean rate $\lambda$.
 \eat{
  
  Consider an M/G/$\infty$ queue in which customers correspond to leechers that own a given tagged block.  The mean residence time of customers in such a queue  corresponds to the mean time that a leecher holds a tagged block, and equals $\sum_{l=0}^{B-1} l/(\mu B) = (B-1)/(2 \mu)$.  The arrival rate of customers is given by a Poisson process with mean rate $\lambda$. }  Therefore, \eat{the probability that that a tagged block is \tunavailable{} among leechers is \eat{number of leechers that own a tagged block is} \eat{given by the num Therefore, the probability that the M/G/$\infty$  queue is empty is} } $q_{\infty}=\exp({-\rho(B-1)/2})$. \eat{, and equals  the probability that a tagged block is \tunavailable{} among leechers.}    In general, a  tagged block is \unsustained{} if no leecher owns the block  and there are no seeds in the system.  The probabilities of these two events are  $\exp({-\rho(B-1)/2})$ and $\exp({-\rho})$, respectively.  Their  product yields~\eqref{esteadyv}. 

\eat{
In~\eqref{esteadyv}, the term ${\nchunks (\nchunks +1)/2}$ corresponds to the fraction of time that a user owns a given tagged block, while the user remains in the system.  Consider a queue that represents the number of users that own a given tagged block.  Since the mean time a user spends in this queue is ${(\nchunks +1)/2}$   The  term $e^{-\rho(\nchunks +1)/2}$  is the probability of 

}

\eat{We now consider the case $\gamma = \infty$.  } 

We  now use the above results to provide a simple expression to estimate the minimum load, $\rho^{\star}$, necessary to attain a given self-sustainability level, $A^{\star}$,  when $\gamma=\infty$. 
It follows from the discussion in the previous paragraph that if $\gamma=\infty$ the probability that a tagged block is \unsustained{}  is  $q_{\infty} =\exp({-\rho(\nchunks -1)/2})$. For values of $q_{\infty}$  close to 0 ($q_{\infty}  \leq 0.01$), $p(\nchunks) \approx 1+E[V]-B=1-B q_{\infty}$, as indicated in the beginning of this section. \eat{ for $\gamma=\mu$.  } This approximation, in turn, can be used to select the \eat{minimum } load  $\rho^{\star}$ to attain self-sustainability level $A^{\star}$,  \eat{ in order to attain a given self-sustainability level, $A^{\star}$, }
\begin{equation} \label{eq:rhostar}
\rho^{\star} \approx \left[2 \log \left(B/ (1-A^{\star}) \right)\right]  / (\nchunks-1), \qquad {\gamma=\infty}
\end{equation}
We further study \eqref{eq:rhostar}  in~\textsection\ref{popularityattainavail}, where we compare the results obtained with this approximation against those  obtained with recursion~\eqref{dimsimple}, that provides exact results. \eat{ In the remark that follows we show that, if we make further assumptions about the regime under consideration, the bandwidth ratios can also be approximated based on the means derived above. }

\subsection{The Impact of File Size on Self-Sustainability} \label{closedform3}

Increasing file size, $B$, \eat{has two consequences: $(i)$ } \eat{increases the number of blocks in the file, $B$, and  }\eat{(which favors a decrease in self-sustainability)}  \eat{$(ii)$} increases the mean download time of peers, $B/\mu$.   \eat{ (which favors an increase in self-sustainability).  } In this section, we show that such an increase in the residence time of peers yields larger \eat{ in impact of the latter on the self-sustainability dominates the , i.e.,} swarm self-sustainability. \eat{ increases with file size.   }  Theorem~\ref{theo:filesize} states the result for \eat{$B$  this is

Although we consider the case } $\gamma=\mu$, $B \ge 4$ and $\rho \ge 1.6$ and  
in~\textsection\ref{sec:evalmain} we provide evidence that  it also holds when $\gamma = \infty$ and for small values of $\rho$. \eat{or $0 < \rho \le 1.6$.}

\begin{theorem}[File Size Impact]  If $B \ge 4$, $\rho \ge 1.6$ and $\gamma=\mu$,  self-sustainability increases with the file size, $B$. \label{theo:filesize}
\end{theorem}

\begin{proof}
We \eat{make the dependence of $\punsust(\nchunks; \mu)$ on $B$ and $\rho$ explicit, and} denote the swarm self-sustainability for a given value of $B$ and $\rho$ as  $\hat{\punsust}(\nchunks,\rho)$.   The Bonferroni inequalities~\cite{bonferroni}, which generalize the inclusion/exclusion principle,   applied to~\eqref{d0}, yield upper and lower bounds on $\hat{\punsust}(\nchunks,\rho)$,
\begin{equation}
1 - B e^{-(B+1) \rho / 2} \leq \hat{\punsust}(\nchunks,\rho) \leq 1 - B e^{-(B+1) \rho / 2} +  {(B (B-1)/2)} e^{-2(B+1) \rho  / 3}
\end{equation}
It is easy to show that if $\rho \geq 1.6$ and $B \ge 4$ then $1 - B e^{-(B+1) \rho / 2} +  {(B (B-1)/2)} e^{-(B+1) \rho 2 / 3} \leq$ $ \qquad 1 - (B+1) e^{-(B+2) \rho / 2} $, from which the result follows by comparing $\hat{\punsust}(\nchunks,\rho)$ and $\hat{\punsust}(\nchunks+1,\rho)$.
{\nobreak\hfill$\Box$}
\end{proof}

The key insight of Theorem~\ref{theo:filesize} can be easily explained  in terms of the busy periods of the proposed model.  The busy period is defined as an uninterrupted interval during which the swarm is self-sustaining.  As the file size increases, the number of blocks that need to be maintained increases linearly but the busy period of the system increases exponentially~\cite{mainarticle}.  Indeed, as the file size increases the availability gain  compensates the overhead to maintain a larger number of blocks, and the self-sustainability increases.  \eat{We further discuss the implications of the file sizes   in~\textsection\ref{popularityattainavail}.}

\section{Evaluation}

\label{sec:evalmain}
In this section we report $(a)$ a validation of the proposed model, against detailed simulations, showing  that despite the  simplifying assumptions considered in our model, it  captures how self-sustainability \eat{and the corresponding  server bandwidth usage } depends on different system parameters and $(b)$ results on the minimum popularity to attain a given self-sustainability level.    \eat{ The goal of the experiments presented in \textsection\ref{sec:parameter} is to}   \eat{  Once our model is validated, in \textsection\ref{popularityattainavail} \eat{and \textsection\ref{sec:numericaleval} } we use it to derive the minimum popularity to attain a given self-sustainability level.} \eat{under varying conditions how publishers should expect the self-sustainability to vary as a function of the load and file size.} \eat{ and chunk distribution policy.}


\eat{Our objectives in this section are (a) to validate the model against
a detailed simulation setup; (b) to compute the performance measures
of previous sections using our model and; (c) to study the influence of
model parameters on the measures.}


\label{sec:exp}

\subsection{Experimental Setup}

Our simulation experiments were conducted using the Tangram-II modeling environment~\cite{tangram09}.   Tangram-II is an event-driven, object oriented modeling tool. \eat{which provides a  convenient framework to obtain the desired metrics of interest.    In Tangram-II } \eat{where actions are triggered by messages transferred between objects.  Tangram-II provides a convenient language  to implement  distributed systems such as BitTorrent. } The three  main objects in our simulations  are the tracker, the peer and the seed.    Their implementations are based on the official BitTorrent protocol description~\cite{bt, legoutrf}.   Every time a peer  enters the system,  receives a block or leaves, we record the event in our logs, the  current timestamp, peer id, and signature  (see ~\textsection\ref{sec:chunkdistrmodel}).~\footnote{
Our Tangram-II model as well as the traces generated for this paper are available at \url{http://www.land.ufrj.br/~arocha/selfsustain}.  }  \eat{For the sake of completeness, we briefly describe our simulator. }

\subsubsection{Simulator and Protocol Descriptions}

\label{simulatordesc}

  When a peer $P$ joins the system,  it receives a random list of fifty other peers from the tracker,  which constitutes its {\em peer set}.   Throughout the simulation, as peers leave the system the size of the peer set of $P$ may dwindle to less than twenty.  Once the peer set size is less than twenty, $P$ requests additional neighbors from the tracker.   The set of peers to whom $P$ offers content blocks is a subset of its peer set, referred to as the {\em active peer set}.

BitTorrent proceeds in rounds of ten seconds.  By the end of each round, peer $P$ runs the tit-for-tat incentive mechanism. \eat{ is executed every ten seconds. } According to this mechanism, $P$ reciprocates contents with those neighbors that contributed in the previous round.   $P$  selects  \eat{uniformly at random} $r$ of those peers that contributed in the last round to add to its active peer set ($r \leq 4$).    In the next round, the  active peer set of $P$ will consist of the $r$  aforementioned peers plus $5-r$ additional peers selected uniformly at random out of its peer set. This random selection of peers is referred to as {\em optimistic unchoke}, performed to allow peers to get bootstrapped as well as to let them learn about new neighbors.   Finally, peers select blocks to download using the rarest first algorithm, except for the first four blocks, which are chosen uniformly.  Each block of the file is divided into sixteen  sub-blocks.  After selecting the blocks to download, peers can get different sub-blocks (of the same block) from different neighbors concurrently.    A block can be uploaded after all its sub-blocks are downloaded, assembled and checked using the block hash key.

\eat{ Those four peers, plus an additional one selected at random, , $\mathcal{P}$ selects  four of them uniformly at random.  In the following round, $\mathcal{P}$ will reciprocate them }  

In our experiments we observe that self-sustainability decreases with the size of the active peer sets. That is because, if the active peer sets are large, each peer splits its bandwidth across many other peers and blocks take longer to get replicated in the network.  As mentioned above, we select an active peer set of size five, which is adopted by many BitTorrent implementations.

\eat{
when the peer enters, he ask random blocks, 4 blocks.  uniform.  256 block, divided into 16 pieces of 16 kb.  priority to finish block as soon as possible. verifica se tem algum bloco incompleto.  10 vizinhos tem bloco, pede um pedaco pra cada vizinho, manda have pra vizinhos. rarest first.   contabilizo os blocos dos meus vizinhos, localmente, 50 peers.  numero de copias, e verifica mais raro.  who has the rarest?  pra nao ficar parado, verifico o segundo mais raro... e assim por diante.  ate alguem que tenha.       }

In \eat{most of } our experiments the seed behaves as a standard peer, except that it $(i)$ initially owns all blocks and $(ii)$ is altruistic, hence does not execute the tit-for-tat algorithm. \eat{ In \textsection\ref{sec:numericaleval} we describe the implementation of a  seed that strategically reveals chunks to the peers.  }
We did   not implement the mechanism used by peers to download their last block, also known as {\em end game mode}~\cite{bt}.  This is inconsequential, though, since the end game mode does not significantly affect the steady state behavior of the system  (see~\textsection\ref{sec:downloadtime} and~\cite{bharambe}).

\subsubsection{Experimental Parameters}

\label{pars}
The configuration of our experiments consists of
torrents that publish a file of size $S$  divided into $B$  blocks of size $s$, $s=256$KB, a typical block size in BitTorrent. The number of blocks in the file, $B$, takes values 16, 50, 100 and 200, which corresponds to files of size 4MB, 12MB, 25MB and 51MB, respectively.  Note that  if a swarm is constituted of multiple files which can be downloaded separately, we are interested in analyzing the self-sustainability of each individual file. \eat{.  Although we initially set out to conduct experiments with larger file sizes, we observed that} A file of size 51MB already yields self-sustainability larger than 0.9 for $\lambda > 0.05$ peers/min (see Figure~\ref{fig:popimp}). Simulations to analyze such a steep increase in self-sustainability (see Figure~\ref{fig:popimp}) \eat{  arrival rates smaller than $0.05/60$ peers/s} quickly requires prohibitively large execution times and significant variability in the metrics of interest across runs. For this reason, we focused on quantitatively validating our model for files with up to 200 blocks, but also use  the model to analyze files of size greater than 200 blocks.

The uplink capacity
of each peer is   $39$KBps, which corresponds to $\mu=39/256=0.15$ blocks/s, a typical effective capacity for BitTorrent peers~\cite[Figure 1]{bittyrant}. \eat{ (which corresponds to a nominal capacity of 
100 KBps in \textsection\ref{planetlab}).  } The 
publisher maximum upload capacity is the same as that of a peer. 
A publisher that contributes the same capacity as an ordinary peer might correspond to either a domestic user \eat{(see~\textsection\ref{sec:motivate})} or a commercial publisher that supports a large catalog of titles and only provides enough capacity for each swarm so as to allow peers to complete their downloads at rate $\mu$.
 The peer
arrival rate is 
varied according to the experimental goals between $(.25, .5, 1, 2, \ldots, 9)$ peers/min, as described
next.

\subsection{Model Validation}

~\label{sec:parameter}

\eat{
We define the publisher bandwidth savings for leveraging peer-to-peer swarming technology, $\beta$, as the ratio of the bandwidth consumed when providing chunks only if they are unavailable among peers over its maximum bandwidth, $\widehat{W} \mathcal{U}$, over its maximum bandwidth, $\widehat{W}$.  
In this section we show experimental and numerical data that provide insight on the impact of different system parameters on publisher bandwidth savings and peers download times.  }  \eat{Our goal now is to show how to parameterize our model in order to capture the publisher bandwidth ratio observed in simulations.  }
Our analytical model makes a number of simplifying assumptions as described in~\textsection\ref{sec:model}. \eat{made in our model include $(a)$ blocks are completely redistributed among peers when the upper layer model transitions from one state to another; $(b)$ the capacity of the system perfectly scales with the number of peers.}  In what follows, our goal is to show that even with those simplifying assumptions, discussed in~\textsection\ref{sec:valiassump}, \eat{ in the light of our simulations, } our model still captures swarm self-sustainability in a realistic setting, \eat{ described in~\textsection\ref{simulatordesc},} as shown in~\textsection\ref{sec:valiesti}.

\subsubsection{Validating Model Assumptions}

\label{sec:valiassump}

\eat{\subsubsection{Distribution and Mean Download Times of Blocks}}
\label{sec:downloadtime}

\begin{figure}[t]
\center
\includegraphics[scale=0.45]{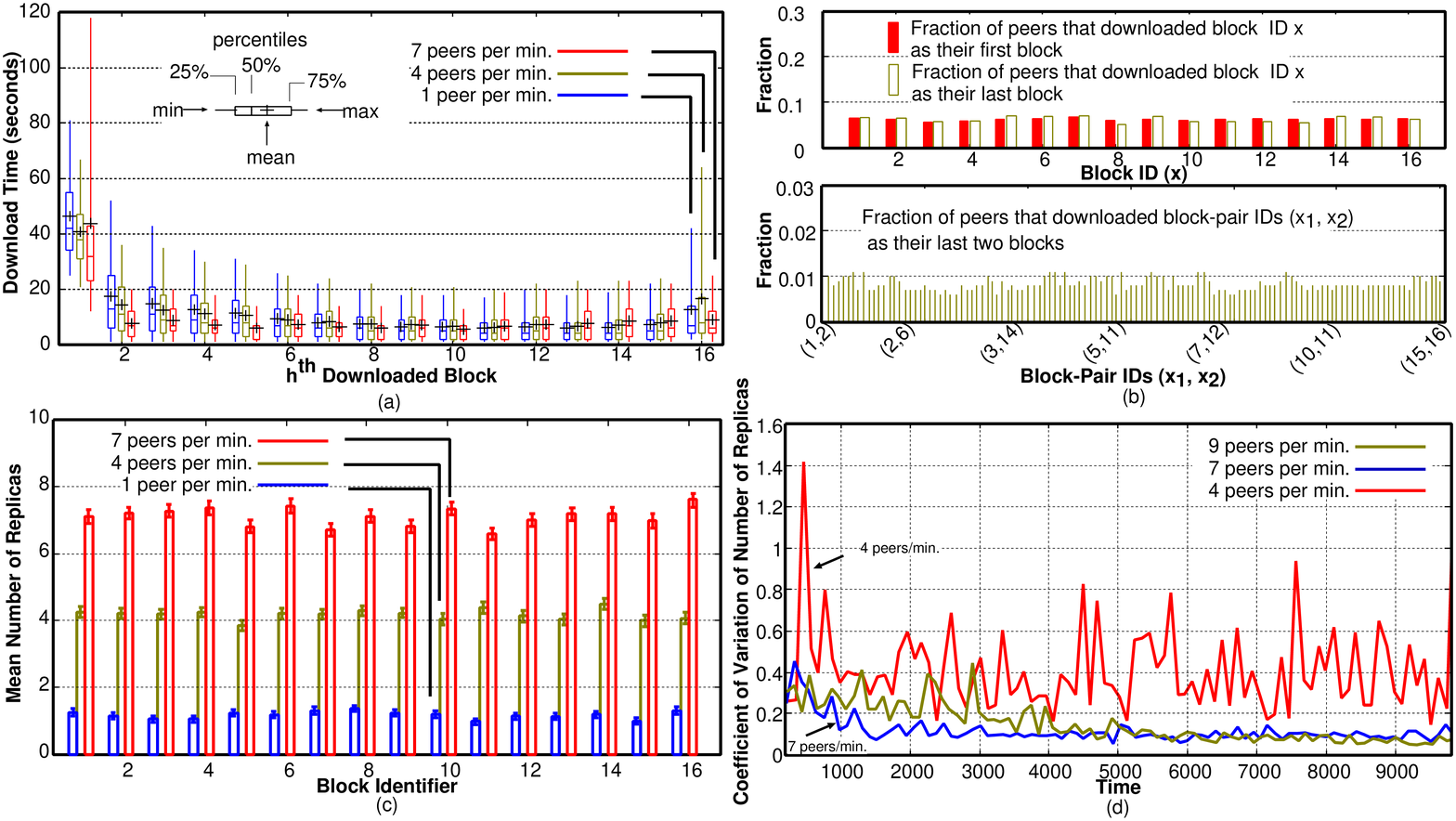}
\caption{$(a)$ Download times of $h^{th}$ block downloaded by a peer.   The boxplots show the four quartiles \eat{The lines show the minimum and maximum values and } and crosses indicate means.    \eat{The download times of the blocks are roughly the same, except for the first and last blocks.  The mean download time of the first block is larger, whereas the last block has a slightly larger mean download time but roughly the same 25th, 50th and 75th percentile.} \eat{ The mean number of replicas of each block is roughly the same. }  $(b)$ Top: shaded (resp., light) rectangles are fractions of peers that download block $x$ as their first (resp., last) block. Bottom: fraction of peers that download block-pair $(x_1,x_2)$ as their last two blocks.  $(c)$ Mean number of replicas of each block with   95\% confidence intervals. $(d)$ Coefficient of variation of number of replicas of blocks versus time. } \label{fig:nreplicas}
\end{figure}

Our aim in this section is to validate $(a)$ that  the mean download times of  blocks  are roughly the same (an exception being the first block downloaded by the peers), i.e., $\mu_h=\mu$, $2 \leq h \leq B$ (see ~\textsection\ref{sec:uplayer}) and $(b)$ that  the signatures of the users are  uniformly distributed (see  \textsection\ref{sec:chunkdistrmodel}).    
  
\eat{We now consider the mean block download times. } \eat{ of blocks. }
\eat{The end game mode does not bring any consequences to our results, since it does not affect the distribution of blocks in steady state. } Figure~\ref{fig:nreplicas}(a) shows the download time of the $h^{th}$ block downloaded by a peer. 
 The boxplots and lines show the distribution  quartiles and minimum and maximum values.  Crosses indicate means download times of blocks, which  are approximately the same, except for the first and last blocks.  
\eat{  The download times of all blocks except the first and the last are roughly the same. } \eat{In particular, note that the download time of the last block is , even without the end game mode being implemented. } In particular, even though our simulator ignores the {\em end game mode}~\cite{bharambe},     in general peers do not experience difficulty finding a neighbor from whom to download their last block. \eat{, which have mean download time sightly higher than the other blocks. } 
  \eat{The maximum download time of the last block is significantly larger than the other blocks, but   the other quartiles } The median of the last block is roughly the same as the one observed for the other blocks, and the mean is only slightly larger.   \eat{That is a consequence of the rarest first algorithm (refer to ~\cite{bharambe} for further discussion on the scenarios under which the end game mode is inconsequential).}  The first block requested by a peer, however, takes longer to be downloaded.  This happens because a peer can only download its first block after being optimistically unchoked (see ~\textsection\ref{simulatordesc}).    Although this affects the time spent by peers in stage zero of our model and as a consequence the total download time, it is inconsequential to our  self-sustainability estimates (the time that peers remain in stage zero of the upper layer model has no influence in our results).  While BitTorrent peers download their first block, they  cannot contribute to self-sustainability as they have no content to provide (see ~\textsection\ref{simulatordesc}).  \eat{ Nevertheless, this is inconsequential to our model parameterization, since in BitTorrent peers can only contribute blocks after concluding the download of the first block (see ~\textsection\ref{simulatordesc}).  The time that peers remain in stage zero of the upper layer model has no influence in our results. }

Our second goal is to study the users' signatures  distribution (\textsection\ref{sec:chunkdistrmodel}).  \eat{Although we initially set out to both validate the uniform assumption and to quantify the extent at which the signatures are correlated, the number of signatures is combinatorial on the number of blocks of the files which makes such analyzes impractical.  } For this purpose, \eat{this reason, } we validated that the first  two and last  two blocks downloaded by a user are indeed uniform and then studied one of the consequences of the uniform and independence assumption, namely, that the number of replicas of each block in the system is well balanced.  

\eat{Henceforth, we let $B=16$.}
Figure~\ref{fig:nreplicas}(b) (top) shows, for each block,  the fraction of peers that downloaded that block as their first block (shaded bars).   \eat{distribution of the identity of the first and last blocks downloaded by users.  }  The figure was generated from independent samples: every  500 seconds, one user owning one block was randomly selected, and the identity of its block was recorded (the same procedure was repeated for users owning all but one block of the file (light bars)).   Similarly, Figure~\ref{fig:nreplicas}(b) (bottom) shows, for every $B(B-1)$ block-pair, the fraction of peers that downloaded that pair as their first two blocks.   Figure~\ref{fig:nreplicas}(b)  indicates that the first and last blocks downloaded by users, as well as the first  downloaded block-pair, are approximately uniformly distributed (the same procedure was repeated for users owning all but two blocks, with similar results).

Figure~\ref{fig:nreplicas}(c) shows the mean number of replicas of each block (including the one stored at the publisher) for $\lambda=1$~peer/min, $4$ peers/min and $7$ peers/min.   \eat{Due  to the rarest-first selection strategy,} The mean number of replicas of each block is around $\rho (\nchunks-1)/2 + 1$, which corresponds to a well balanced system (see~\textsection\ref{sec:minloadanl}).   Figure~\ref{fig:nreplicas}(d) corroborates this claim by showing the coefficient of variation of the number of replicas of blocks \eat{(standard deviation of number of replicas divided by mean number of replicas)} as a function of time.   Let $r_{i,t}$ be the number of replicas of block $i$ at time~$t$.  The mean number of replicas of blocks at time~$t$ is $\mu_t =  \sum_{i=1}^{B} r_{i,t}/B$ and the coefficient of variation  is $c_t=\sqrt{(\sum_{i=1}^{B} (r_{i,t} - \mu_t)^2 / B)}  /\mu_t$. Figure~\ref{fig:nreplicas}(d) indicates that throughout the simulation, the coefficient of variation is usually smaller than 0.8, which means that  the number of replicas of blocks has a low variance. In a system where users  signatures  are uniform and independent we would observe similar behavior.  \eat{Even though in Figure~\ref{fig:nreplicas}(a) we plot only steady state metrics, we observed that the number of replicas of blocks was well balanced throughout our experiments.  } \eat{ Another possibility for the uniform chunk distribution model to match the simulations so well as  may be be that our ideal rarest first distribution model is too optimistic.   We leave this question open for future investigations,  but noting that the ideal rarest first and the uniform distribution policy are abstractions about two extremes scenarios. \eat{, which we further compare in~\textsection\ref{sec:numericaleval}.} }

\eat{
 The end game mode may help peers to download the last chunk faster.  1eless, it is inconsequential to our results, since it does not affect the distribution of chunks in steady state:  $(a)$ if the last chunk is highly replicated, a peer will be able to download it regardless of the end game mode; $(b)$ if the chunk is difficult to find, a peer will not take advantage of the end game mode, since there will be no multiple sources from whom to download  the chunk.    In any case, peers leave the system after downloading the last chunk, which implies that downloading it faster will not affect content availability.   }

\subsubsection{Validating Model Estimate of Self-Sustainability}

\eat{\subsubsection{The Impact of Content Popularity}}

\label{sec:valiesti}

To study how content popularity impacts self-sustainability,  we simulated BitTorrent in the setting described in~\textsection\ref{pars}, varying the arrival rate of peers, $\lambda$, from 1 peer/minute  to 8 peers/minute, in increments of 1 peer/minute,  while maintaining all other parameters fixed.  Equivalently, this corresponds to an increase in the load, $\rho=\lambda s/{\mu}$, from 0.1 to 0.8.    Each simulation lasted for 10,000s. Twenty one  independent simulations were executed for each value of $\lambda$, and used to compute 95\% confidence intervals.  The same experiment is repeated for $B=16$, $50$, $100$ and $200$.  

\begin{figure}[t] 
\center 
\includegraphics[scale=0.55]{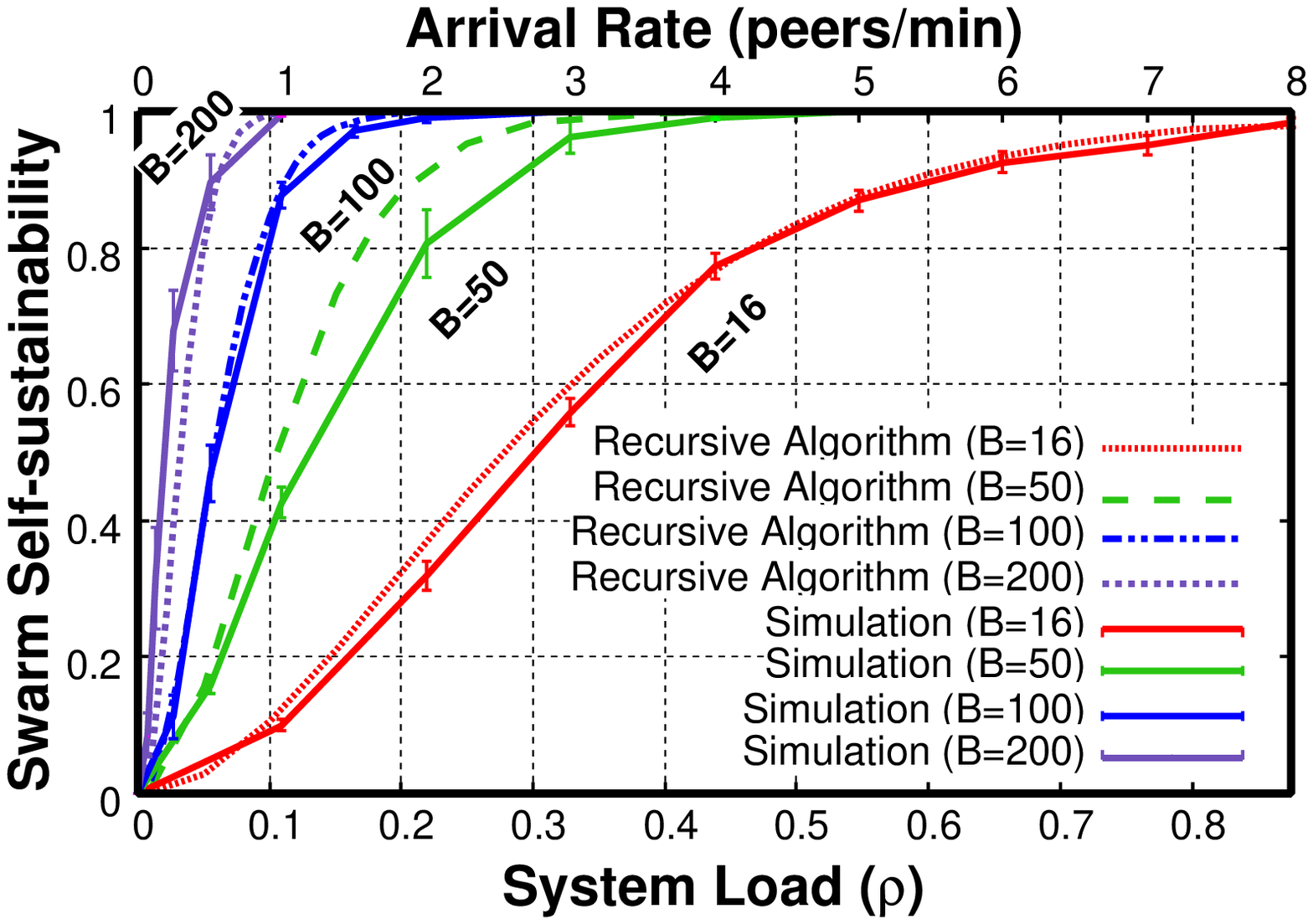} 
\caption{Model validation. Swarm self sustainability as a function of the system load.     Results obtained with recursive algorithm and simulations (with 95\% confidence intervals) are plotted with dotted and solid lines, respectively.       } \label{fig:popimp}
\end{figure}

Figure~\ref{fig:popimp}  shows self-sustainability, $A$, as a function of the content popularity, $\lambda$, for $B=16, 50, 100$ and $200$.  \eat{The lowest curve is obtained from our  simulations. }  For unpopular contents, $\lambda =  1$ peer/min, and small files, $B=16$, swarm self-sustainability is around 0.1 and the publisher needs to frequently provide blocks that are \unsustained. \eat{ and the bandwidth ratio is close to 1.  } As the popularity of the files increases, swarm self-sustainability increases and content is available even in the absence of the publisher.  For $\lambda = 8$ peers/min, the fraction of time at which the publisher needs to provide blocks to peers is close to zero.

 Figure~\ref{fig:popimp} indicates that the results obtained with our  model are close to those obtained through simulation.  Even \eat{Although we used our model in its simplest possible form, e.g., } assuming that the mean download times of all blocks are the same ($\mu_h = \mu$, for $1 \leq h \leq B$), the model was  able to capture the self-sustainability observed in our simulations.     We also repeated the simulations with heterogeneous peer upload capacities, for $B=16$ (the upload rate distribution taken from the measured data used to generate Figure 1 in the BitTyrant study~\cite{bittyrant}) and our results did not qualitatively change (details in~\cite{technicalreportq}).  \eat{ In particular, the self-sustainability predicted by the model always falls within the confidence interval of the simulation.} \eat{\m\punsust_i=\mu$, $0 \leq i \leq \nchunks -1$. } \eat{ Using  the uniform chunk distribution policy at the lower layer, }

\eat{ \textbf{Prototype Experiments.} We now briefly indicate that our results are consistent with the ones  reported by~\cite{mainarticle}.  In~\cite{mainarticle} the authors present results  using PlanetLab experimetns under the same reference setting as the one described in~\textsection\ref{pars}, with $B=16$. The nominal peer capacity of 100KBps  used in~\cite{mainarticle} corresponds to an effective capacity of 39KBps.   
 As shown in Figure~\ref{fig:popimp} and noted above, when $\lambda > 7/60$~peers/s the swarm is almost always self-sustaining. This means that peers can download content for a long time even in the absence of a publisher, an observation that is consistent with the ones made in~\cite{mainarticle}. 

}

\eat{

Our model describes how the steady state chunk unavailability increases as a function of the load in the system.  Recall that the load is defined as $\lambda s/\mu$.   For a fixed file size and peer capacity, as the popularity of the file increases the chunk availability among the peers increases 

}


\eat{
Figure \ref{fig:model-validation} plots the chunck availability among peers
versus the system load for several buffer size values.   
Also in the figure we plot the results from the detailed simulation for
$B=16$.
>From the figure it is evident that the uniform chunck distribution assumption
closely matches the simulation results.
}

\eat{The predictions of our model for  $\nchunks=100$ are also plotted in Figure~\ref{fig:popimp}. }\eat{ also shows the predictions of our model for . }

\eat{\subsubsection{Impact of the File Size}}

Consider now the impact of file size on swarm self-sustainability. Figure~\ref{fig:popimp} shows that 
for a fixed content popularity, as file size increases,  self-sustainability increases.  This is in accordance with Theorem~\ref{theo:filesize}, and reflects the fact that, as file size increases, peers stay longer in the system and the coverage, defined as the mean number of users in the system, increases.  The higher the coverage, the greater the self-sustainability of the swarm.  In fact, as file size increases, the number of blocks that needs to be maintained by the publisher increases linearly but the availability gain increases exponentially~\cite{mainarticle}. \eat{.  This property can be explored by the publisher when deciding how to bundle its files}

\eat{
We observe the sharp increase on chunck availability on the content popularity.
For instance, for file sizes of $4$ MB, doubling the load from $0.3$ to $0.6$
increases the availability from $0.54$ to $0.91$, that is, the expected
long range bandwidth $W$ decreases by approximately 80\%  

Likewise, if the load increases from $0.05$ to $0.15$ the bandwidth savings is
96\%.  (that is, of the popularity increases from $0.46$ peers per minute to $1.4$.
}

\eat{
*** {\bf we define bandwidth savings as ( (1-Av2) - (1- Av1) )/ (1-Av2), were
Av2 < Av1.  Perhaps we should plot the bandwidth savings, not the Av).} }

\eat{ Our model suggests that ... as discussed above. ...}

\subsection{Popularity to Attain  High Self-Sustainability}

\label{sec:numericaleval}

\label{popularityattainavail}

\begin{figure}[t]
\center
\includegraphics[scale=0.4]{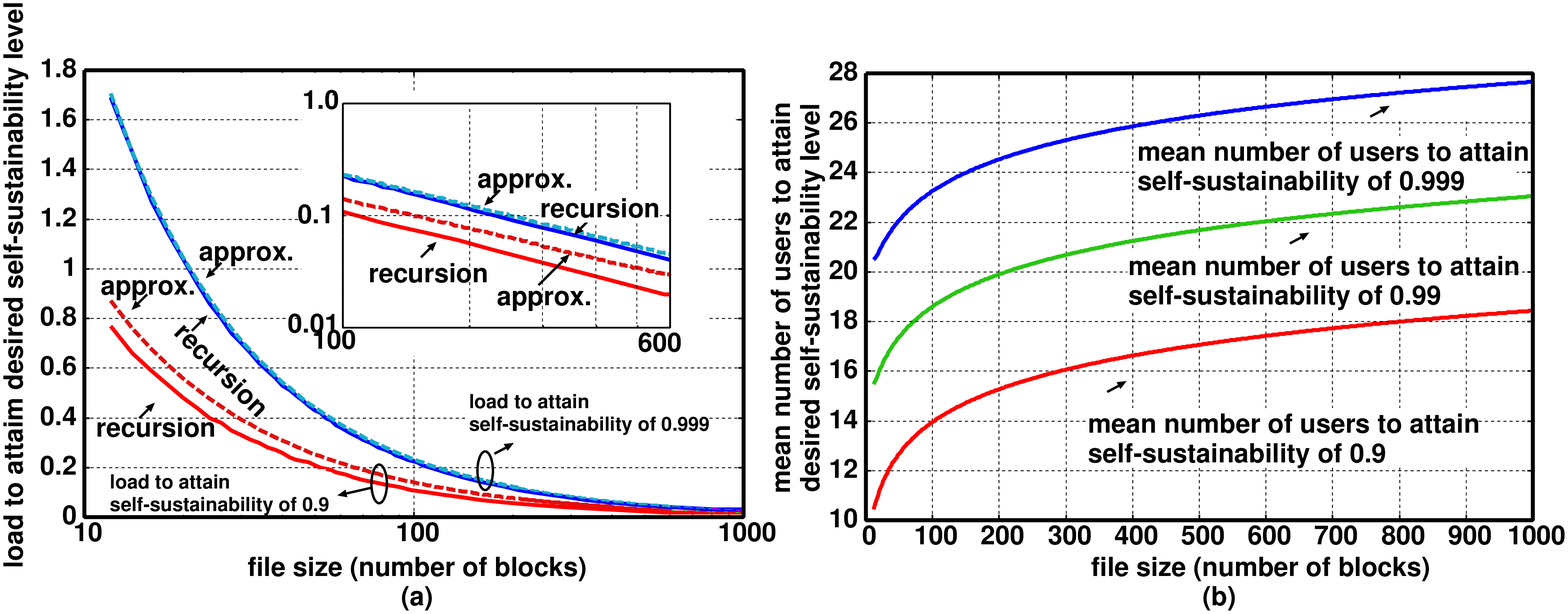} 
\caption{Larger files yield increased availability. $(a)$  $x$-axis, file size.  $y$-axis, necessary load ($\rho^{\star} =\lambda/\mu$) to attain self-sustainability greater than 0.9 (red) and 0.999 (blue). The results obtained using the approximation~\eqref{eq:rhostar} are also shown.  \eat{Part (a) zoom:  $\log \rho^{\star}$ decreases linearly in $\log B$.  } $(b)$  the mean number of users in the system, $B\rho^{\star}$,   to attain a desired self-sustainability level.    } \label{fig:fig-bounds}
\end{figure}

We now address the following question: what is the minimum file popularity  (or load) needed to attain a given self-sustainability?  \eat{Recall that the {\em load} is defined as the popularity of the file times the mean time for peers to download a block, ${\rho}=\lambda/\mu$.   Our question can be restated in terms of the load as: what is the minimum load in the system so as to achieve the desired block availability and corresponding publisher bandwidth usage?  } Answering this question is useful not only for publisher dimensioning but also for other strategic decisions such as how to distribute and bundle  files across multiple swarms~\cite{mainarticle}.

 \eat{Let $\epsilon$ be the fraction of time in which at least one block is \unsustained{} and  $1-\epsilon$ the corresponding self-sustainability of the swarm.}

 Recall that the load is defined as  $\rho=\lambda /\mu$. 
Figure~\ref{fig:fig-bounds} shows the minimum load, $\rho^{\star}$,  necessary to achieve high self-sustainability, $A^{\star}$ ($A^{\star}=0.9,0.999$), for file sizes  varying between 2MB and 256MB ($B=8, \ldots, 1,000$).   This figure illustrates self-sustainability as predicted by the recursions contained in Theorem~\ref{theo:efficientalgo}, eq.~\eqref{dimsimple}  \eat{,  selecting the appropriate $\rho$ to achieve chunck 
availability among peers above $1-\epsilon$. } and the approximation~\eqref{eq:rhostar}.  
Figure~\ref{fig:fig-bounds}(a) suggests that when the goal is to find the minimum popularity to attain a high self-sustainability level,  
equation~\eqref{eq:rhostar}  can be used \eat{the shape of the function} to approximate $\rho^\star$.  However, if the goal is to compute self-sustainability under different loads and for different files sizes, as illustrated by Figure~\ref{fig:popimp}, the recursion provided by Theorem~\ref{theo:efficientalgo} needs to be used.

\eat{
plots, for different values
of file sizes, the load needed to achieve a high level of
chunck availability among peers or, equivalently, the minimum popularity
threshold values such that the publisher bandwidth ratio is very low
(say below $10^{-3})$.

File sizes range from 2 MB (8 blocks) to $2.56$ GB (10.000 blocks).
The block distribution is assumed uniform.
}

\eat{
In Figure \ref{fig:fig-bounds} two curves are shown:
the red curve is obtained from the exact recursive algorithm
(by selecting the appropriate load value to achieve a high chunck 
availability among peers);
and the blue curve from equation~\eqref{eq:rhostar}.}

\eat{This is explained as follows.  
  When the mean fraction of unavailable blocks, $p$, is small, it constitutes a good approximation for the probability of no blocks being unavailable, $P(V=0)$.  Therefore, given the  target bandwidth ratio, $\epsilon$, we can use~\eqref{eq:rhostar} to compute $\rho^{\star}$ as a function of $p \approx \epsilon$ and $B$.   According to~\eqref{eq:rhostar}, for a fixed $\epsilon$ we have that  $\log \rho^\star$ is  linear in $\log B$. }  \eat{
This is exactly what one would expect if we use equation~\eqref{eq:rhostar},
and replace the mean fraction of unavailable chucks, p, by the probability
that no chuncks are missing among peers, . }

\eat{
Approximating the probability of no blocks being unavailable by the fraction of unavailable blocks is motivated by Lemma~\ref{theoappr}.  
  For very large  files
(for example, $B \ge 5,000$), and for a fixed upper layer state,  Lemma~\ref{theoappr} shows that the 
fraction of unavailable chuncks
is concentrated around its mean. }
\eat{
When the file size is very high (for example, $B \ge 5,000$) 

Given the upper layer state $\bncustomers$,  Lemma~\ref{theoappr} shows that the 
fraction of unavailable chuncks
is concentrated around its mean when the number of chuncks is very high
(for example, $B \ge 5,000$).
}
\eat{
So, we would expected that equation~\eqref{eq:rhostar}
should be used to computed the probability of zero chuncks missing among the
peers only in these cases. }  \eat{
Nevertheless, the exact results in }  \eat{constitutes is a good approximatversus
the chunck size is preserved } \eat{as a pretty good approximation, } \eat{even for relatively small files.} \eat{chunck sizes.}  \eat{This  simple relationship between    is remarkable. }  \eat{, and
it can be used to estimate the minimum file popularity needed to
attain a high level of chunck availability (or very low
publisher bandwidth usage). }

Figure~\ref{fig:fig-bounds}(a) indicates that the popularity, $\rho^\star$, needed 
to attain a high degree of self-sustainability increases as the
file size, $B$, decreases.  In particular, the zoom in the figure shows that $\log \rho^\star$ is  linear in $\log B$.    The comments made at the end of ~\textsection\ref{sec:parameter}  to explain Figure~\ref{fig:popimp} also apply here.  Peers take longer to download larger files, which increases block availability.     Nevertheless, the benefits of leveraging peer-to-peer swarming can be noted even for small files. Figure~\ref{fig:fig-bounds}(a) shows that for a file of $4$MB ($B=16$), an arrival rate of $8$ peers/min (which corresponds to a load of 0.8) already yields a very high self-sustainability.

More insights on how file size impacts self-sustainability are obtained from Figure~\ref{fig:fig-bounds}(b).   Figure~\ref{fig:fig-bounds}(b)  plots the   mean number of users in the system, also referred to as the mean coverage~\cite{mainarticle}, necessary to attain a high level of self-sustainability.    The curves in Figure~\ref{fig:fig-bounds}(b) \eat{corresponding to a self-sustainability of 0.9 and 0.999} correspond to the the respective ones in Figure~\ref{fig:fig-bounds}(a)  multiplied by $B$.  The main insight shown in this example is that the  coverage necessary to attain a given self-sustainability level slowly increases as a function of $B$.    As the file size increases, a slightly larger population suffices to attain a given  self-sustainability level.    
  For instance, for a file of size 10 a coverage  of 20 is necessary to attain  self-sustainability of 0.999, whereas a coverage of 29 suffices to achieve the same self-sustainability if the file has 1,000 blocks.

\eat{This is intuitive because the As peeless time to download small files
in comparison with large ones. }   
\eat{Figure~\ref{fig:fig-bounds}  shows that even in the absence of a publisher, peers can almost always one can maintain a considerably peer bandwidth
savings (or, in other words, one can maintain the swarm with little help from the seeder)
even for small file sizes (say $8$ chuncks or $2$ MB) with a reasonable
file popularity (23 peers per minute in this example). }

\eat{\subsection{The Impact of Altruistic Peers}}

\section{Discussion}

\label{assumptions}

\eat{ \eat{simplicity, } the design of our model was guided by two general principles: usefulness and simplicity.} \eat{ \eat{ were followed while devising the proposed model.  } From a pragmatic standpoint, a model is good if it is useful.  In~\textsection\ref{sec:evalmain} we validated the applicability of our model against simulation experiments, showing that it captures the experimental results with high accuracy.   From a mathematical standpoint,  a model should be as simple as possible.  } Next, we discuss the assumptions adopted \eat{in our~model } to yield a tractable model.

{\bf{Uniformity and independence assumptions: }}    In the performance model presented in~\textsection\ref{sec:lowlayer} we assume that the signatures of users are drawn uniformly and independently at random.  In particular, we do not account  for correlations \eat{ of users' signatures over time or} among users' signatures.  \eat{These simplifications might be too stringent, for instance, in order to model flash crowds. However, } Although such correlations are present in practice, our simulations have indicated that the independence  assumption is  appropriate in order to capture the self-sustainability of  swarms.  \eat{A consequence of the uniformity and independence assumptions is that the number of replicas of each block is well balanced, a fact that we validated with our simulations.   }  \eat{ satisfying the steady state assumption. }  \eat{As a side note, we alternatively  considered a performance model according to which a block is missing if and only if the users do not, collectively, own $B$ replicas of blocks of the file.  However, such model turned out to capture an idealized scenario which did not correspond to the one observed in our simulations and for this reason we omit its description in this paper.}

In the user dynamic model presented in~\textsection\ref{sec:uplayer} we make the following  assumptions: $(i)$ peers arrive according to a Poisson process with rate $\lambda$  (steady state assumption),  $(ii)$ the download times of all blocks have the same mean, $1/\mu$ (smooth download assumption) and $(iii)$ users leave the system immediately after completing their downloads, $\gamma=\infty$ (self-regarding users assumption).  We discuss each of these in turn.

{\bf{Steady state assumption: }} It has been shown in ~\cite[Section 4.3.4]{mainarticle} that a vast number of long-lived swarms have relatively stable mean arrival rates over periods of months.   Our model can be used to predict the self-sustainability of such swarms.   \eat{ In some cases, the model may also be used to lower bound the self-sustainability of swarms with non-stationary arrival rates.}  

\eat{
{\bf{Homogeneous population assumption: }} We assume that all peers, including the  publisher, contribute with an upload rate of $\mu$.   Note that a publisher that contributes with the same rate as ordinary peers might correspond to either a domestic user \eat{(see~\textsection\ref{sec:motivate})} or a commercial publisher that supports a large catalog of titles and only provides enough capacity for each swarm so as to allow peers to complete their downloads at rate $\mu$.
}

{\bf{Smooth download assumption: }} Under this assumption, the capacity of the system scales perfectly  with the number of users.   Our simulations indicate  that the mean download time of the first and last blocks are slightly larger than the others.  Although  our model has the flexibility to capture such asymmetries (see observation two in~\textsection\ref{sec:bwratio}), we show that  their implications in the estimates of  self-sustainability are  not significant \eat{for our purposes} (see~\textsection\ref{sec:downloadtime}).

{\bf{Self-regarding users assumption: }} \eat{means that users depart immediately from the network after completing their downloads.   } Our model has the flexibility to account for users that stay in the network after completing their downloads. However, in today's BitTorrent users have no incentive to stay in the system after obtaining the files of their interest.  Therefore,  we focus on the worst case scenario in which users, not having incentives to linger in the system after completing their downloads, depart immediately.

Finally, in our simulations we consider a publisher that is always online and that behaves like a typical peer.

{\bf{Typical peer-like publisher: }}  Our simulations indicate that if the publisher has the same capacity as a typical peer, the smooth download assumption holds and the swarm self-sustainability estimates of our model are accurate. Coping with intermittent publishers and devising dynamic bandwidth  allocation strategies according to which the smooth  download assumption holds  is non trivial~\cite{syndrome}, and is subject of future work.

\section{Related Work and Discussion}

\label{sec:related}

\subsubsection*{Modeling of Peer-to-Peer Swarming Systems}

The literature on availability~\cite{neglia}, performance~\cite{qiusrikant} and incentive issues~\cite{bittyrant} in BitTorrent-like swarming systems is vast.  Nevertheless, to our knowledge this paper presents the first analytical model for publisher dependency estimation. \eat{ accounting for the distribution of chunks in the swarm.  } \eat{More generally, } We are unaware of related analytical work that analyzed swarm self-sustainability \eat{, in particular}  \eat{steady state availability of content among peers  in swarming systems}  taking into account the fact that files are divided into multiple blocks, a very fundamental characteristic of these systems.  

For large populations, Massouli\'{e} and Vojnovic~\cite{laurent} used a 
coupon collector model to show that asymptotically the distribution of blocks across the population is well balanced, and does not critically depend on the block selection algorithm used by the peers. \eat{ leading to a
robust system. } Qiu and Srikant~\cite{qiusrikant} and Fan et al.~\cite{fanchiului} also   considered the large
population regime, and used  fluid approximations and 
differential equations to model the system assuming that the efficiency  is always 
high.   In this paper we are particularly interested in the small population regime.  For small populations, Markov Chain (MC) models have been proposed by Yang and Veciana~\cite{veciana}, providing insights on the performance of the system but not dealing with the problem of block availability among peers.  \eat{availability.  }     Norros and Reittu~\cite{toward}, using a different model,  studied the dissemination of a two-block file in a closed network accounting for the availability of the  blocks.   In this paper, we consider an open network and propose a model which can be used to estimate self-sustainability of files of arbitrary size.

\eat{
Using a model different from ours,  Reittu et al~\cite{toward}   The authors consider a file that is divided into two blocks. Initially the two blocks are available but after some iterations blocks may become unavailable.  The authors aim to find the steady state  distribution of the number of available blocks. In this paper, we propose a model which can be used to analyze files of arbitrary size. }

In this work we studied the implications of  the content popularity on the self-sustainability of swarms.\eat{arrival rate of peers. } \eat{there is a phase transition on the availability as a function of the arrival rate.  Phase transitions in  peer-to-peer systems have been observed by other authors. } In a real time setting, Leskela et al.~\cite{stability} pointed out a phase transition in the stability of the peer-to-peer system as a function of the content popularity.  In contrast, our system is always stable.  Norros et al.~\cite{iwqos07} imply a phase transition of the mean broadcast times as a function of the departure rate of seeds.  In this paper we are concerned with self-sustainability.   \eat{ The last authors also point out that  in a system in which  peer encounters are random the distribution of chunks is uniform~\cite{flash, iwqos07}.  }

\subsubsection*{Hybrid Peer-to-Peer Swarming Systems}

The literature on the use of peer-to-peer swarming systems for enterprise content delivery is rapidly growing ~\cite{antfarm, serversp2p, dimensioning}.  \eat{This reflects the fact that peer-to-peer systems can potentially benefit publishers, but their widespread use is still to happen.  } The  methodology usually consists of defining an optimization problem to be solved by the publishers and then showing how different system parameters affect the optimal bandwidth allocation strategy.  The approach we take in this paper is different.  We are interested in the \emph{ minimum } fraction of time that the publisher must be active so as to guarantee that all blocks are always available. \eat{From the optimization perspective, this corresponds to a lower bound on the  bandwidth allocated by the publisher that must be spent if content availability has to be ensured.  } 

Ioannidis and Marbach~\cite{ioannidis} study how quickly the bandwidth available at the server has to grow as the number of users increases.   For this purpose they consider two query propagation mechanisms, the random walk and the expanding ring.  Here, on the other hand, we assume that peers can always find the blocks they need in case they are available.   While ~\cite{ioannidis}  focuses on the control plane and it's asymptotic analysis, here we focus on the data plane and account also for small files.  

\eat{Estimating the minimum bandwidth used by the server is closely related to  estimating  chunk availability among peers.  }

Menasche et al.~\cite{mainarticle}, Wong et al.~\cite{altman}  and Susitaival et al. ~\cite{susitaival} propose models for content availability in BitTorrent without accounting for the fact that files are divided into blocks.    Our model differs from ~\cite{mainarticle, altman, susitaival} in that we $(a)$  consider a hybrid peer-to-peer system, in which a publisher is always available and $(b)$ account for the fact that the file is divided into blocks.     \eat{ and $(c)$ capture the implications of different chunk dissemination policies.} \eat{ a stable publisher to provide chunks quantifies content availability at the chunk level while prior work did not account for the fact that files are divided into chunks. } \eat{quantifies availability at the content level, assuming each file is composed of one single chunk. } \eat{To the best of our knowledge, we are the first to propose a model for availability accounting for the course grained division of a file into chunks.  } Finally, explicit scheduling of blocks exchanges to minimize peer download times was studied by Mundinger et al.~\cite{mundinger}.   In this paper we assume that peers exchange blocks using only local information, as in BitTorrent.

\subsubsection*{Balls and Bins}

The derivation of some of our results fit into the balls and bins framework.   Each user is allocated a  set of blocks (balls) each of which must correspond to a different identifier (bin).  In the context  of balls and bins, a set of balls each of which must be allocated in a different bin is referred to as complex~\cite{randomallocation}. Previous work on the allocation of complexes into bins appears in  Mirakhmedov and Mirakhmedov~\cite{saidbek}, Kolchin et al.~\cite[Chapter VII]{randomallocation}, and references therein.   In particular, the definition of $\psi_h(i,m)$ in this paper was inspired by~\cite[Figure 1]{balanced}.    

In a peer-to-peer setting, balls and bins were used by Simatos et al.~\cite{simatos} to study the duration of the regime during which the system is saturated because  capacity is smaller than  demand.  The scenario studied  in this paper differs from ~\cite{simatos} in several aspects. For instance, ~\cite{simatos} considers a finite population of peers.

\section{Conclusion}

\label{sec:conclusion}

\eat{
Even though  peer-to-peer swarming systems are well suited to assist  content delivery, there are still a number of challenges that need to be faced before their widespread use. 

}

Peer-to-peer swarming systems are a powerful tool for content delivery, as reflected by the immense popularity of BitTorrent and the vast literature on the topic.   However, most works in this area have focused on the dissemination of popular content, for which peer-to-peer systems are naturally suitable.  In this work, we investigate the dependency of peers on a publisher that leverages peer-to-peer techniques for the dissemination of both popular and unpopular content.  In particular, the latter deserve special attention, since unpopular content can represent a significant fraction of demand and revenue~\cite{anderson}.  We believe that devising strategies for disseminating large catalogs of files leveraging peer-to-peer techniques is an important and interesting research area, and we  see our model as a first attempt to shed light into  the intrinsic advantages and  limitations of peer-to-peer swarming systems for the dissemination of such catalogs.

\bigskip

 \textbf{Acknowledgements} 
We are thankful to  Carsten Schneider, author of Sigma~\cite{sigma}, for his support using the package.  This work was supported in part by the NSF under award numbers  CNS-0519922 and CNS-0721779.  Research of DSM  also sponsored by a scholarship from CAPES/Fulbright (Brazil).   Research of ESS, RML and AAAR is partially funded by research grants
from CNPq and FAPERJ.





\bibliographystyle{elsarticle-num}
\bibliography{selfsust-els}

\begin{appendix}

\section{An Expression of  $P(V=\nchunks|\bNCUSTOMERS=\bncustomers)$}

\label{sec:apppv}

We now show how to obtain $P(V=\nchunks|\bNCUSTOMERS=\bncustomers)$  from~\eqref{uniformjoint}.  Recall that, given an upper layer state $\bncustomers$,  $\Omega_{\bncustomers}$ is the lower layer sample space.   Denote by $\bS$ the random variable that represents the lower layer state, and by $\bs$ its realization.  Then
\begin{equation}
P(V=\nchunks|\bNCUSTOMERS=\bncustomers)= \sum_{\bs:V=\nchunks, \bs\in \Omega_{\bncustomers}} P(\bS = \bs | \bNCUSTOMERS = \bncustomers) \stackrel{\eqref{uniformjoint}}{=} \sum_{\bs:V=\nchunks, \bs\in \Omega_{\bncustomers}} 1/|\Omega_{\bncustomers}|
\end{equation}
Therefore, the problem of computing  $P(V=\nchunks|\bNCUSTOMERS=\bncustomers)$ is reduced to that of counting the states in which all blocks are \sustained.  Using the inclusion/exclusion principle,
$P(\unavailable = \nchunks|\bNCUSTOMERS = \bncustomers) =   {|\Omega_{\bncustomers}|}^{-1} \sum_{i=0}^{\nchunks-1} (-1)^{i} {\nchunks \choose i}  \prod_{j=1}^{\nchunks-1} {{\nchunks-i \choose j}}^{\ncustomers_j} .$ 
In general, $P(\unavailable = \realav|\bNCUSTOMERS = \bncustomers)$ is also obtained using the inclusion/exclusion principle~\cite[Section 4.2]{genera}.

\section{Recursion to Compute $\psi_h(k,\realav)$}

\label{sec:apprecpsi}

Consider the scenario  where  $k$ blocks are \sustained{} and an additional user contributes  $h$ blocks.   $\psi_h(k,\realav)$ is the probability that $\realav$ blocks are \sustained{} after accounting for the blocks contributed by the additional user.
$\psi_h(k,\realav)$ can be recursively computed, 
\begin{equation} \label{recpsi} 
\psi_h(k,\realav) = \psi_{h-1}(k,\realav-1) \frac{B{-}\realav{-}h}{B{-}h{+}1} + \psi_{h-1}(k,\realav) \frac{\realav}{B{-}h{+}1}, \qquad 0 \leq k, \realav \leq B, \quad 0 < h \leq B  
\end{equation}
The base cases are $\psi_{0}(k,\realav)=0$ if $\realav \neq k$ and $\psi_{0}(k,\realav)=1$ if $\realav = k$.  Note that the above recursion is convenient to avoid numerical problems since it only involves additions and  multiplications of probabilities.

   For presentation convenience, we consider an arbitrary ordering of the $h$ blocks contributed by the additional user.  After contributing the first $h-1$ blocks, there are two cases to consider, $(i)$ $\realav$ blocks are \sustained{}  [and the  $h^{th}$ block overlaps with a previously \tavailable{} block, an event which  happens with probability  $\realav/({B{-}h{+}1})$] or $(ii)$ $\realav-1$ blocks are available [and the $h^{th}$ block does not overlap with previously available blocks, an event which  happens with probability $(B{-}\realav{-}h)/({B{-}h{+}1})$].   Cases $(i)$ and $(ii)$ correspond to the first and second terms in~\eqref{recpsi}, respectively.
\eat{For the $h < 2$ or $\realav=0$, we use~\eqref{psiimh}. }

\section{Proof of Lemma~\ref{lemmaprob}}

\label{sec:applemma}

We now show that~\eqref{mainrec}  and~\eqref{umpvm} are equivalent. To this purpose, we consider an extended representation of the upper layer states.  In such representation, \eat{when there are $|\bncustomers|$ users in the system, } the upper layer state, $\bncustomers'$, is \eat{given by two vectors.  Vector $\bncustomers$ was defined in ~\textsection{\ref{sec:uplayer}}, and $\bncustomers'$ is } defined as follows.

  Consider the users in the system ordered uniformly at random. The number of blocks owned by the $i^{th}$ user is denoted by $\ncustomers'_i$. The  upper layer state is represented by vector  $\bncustomers'$.  \eat{We denote the number of users in the system when the state is $\bncustomers'$ by $|\bncustomers'|$.   } The dimension of $\bncustomers'$  is $1\times n$, where $n$ is the number of users in the system.   
  
  Note that $\bncustomers$ is inferred from $\bncustomers'$. Let $n_l(\bncustomers')$ be the number of users that have $l$ blocks when the  state is $\bncustomers'$.  \eat{, we characterize the upper layer  state by $(\bncustomers,\bncustomers')$  for convenience, which allows us to refer to $\bncustomers$ or $\bncustomers'$ as needed. } Since any permutation of the users in the system is equally likely, the steady state probability of state $\bncustomers'$ is $\pi(\bncustomers')$, 
 $$ 
\pi(\bncustomers') =   \prod_{l=0}^{\nchunks-1}  \left[     \frac{\rho^{n_l(\bncustomers')}}{n_l(\bncustomers')!} e^{-\rho}  \right] \frac{1}{n!}
$$
 \eat{This, in turn, follows from the memoryless property of the exponential distribution and the fact that we order users based on the time it takes for them to complete the download of their corresponding next blocks. } The steady state probability of state $\bncustomers'$, conditioned on the event that there are $n$ users in the system,   is 
 \begin{equation} \displaystyle
\frac{\pi(\bncustomers')}{\sum_{\forall \bncustomers': |\bncustomers'|=n} \pi(\bncustomers')}  {=}  \frac{ \pi(\bncustomers')}{e^{-B \rho} (B\rho)^{n} / {n}!}
\label{condprobv}
=   \frac{1      }{ B^{n} \prod_{l=0}^{\nchunks-1}  {n_l(\bncustomers')!}}
\end{equation}
The first equality follows from the fact that a superposition of $B$ Poisson processes with rate $\rho$ is a Poisson process with rate $B \rho$.

The key idea of the proof consists of partitioning the state space  into sets $G_h(n)$, $0 \leq h \leq \nchunks-1$, $0 \leq n$.  Set $G_h(n)$ contains states in which there are $n$ users in the system and \eat{the next user to complete the download of a block currently owns $h$ blocks.} $\ncustomers'_n=h$.  $G_h(n)$ is defined as 
 $G_h(n) = \{ \bncustomers' : \ncustomers'_n = h \textrm{ and } 0\le \ncustomers'_j \le B-1, 1 \leq j < n \}$. 
The set  containing all states in which there are $n$ users in the system, $G(n)$, is 	

\begin{equation} 
G(n) = \cup_{h=0}^{B-1} G_h(n) \label{unionb} = \{ \bncustomers' : 0 \le \ncustomers'_j \le B-1, 1 \leq j \leq n  \} 
\end{equation}

Note that $G_h(n)$ is obtained from $G(n-1)$ by adding to each element of $G(n-1)$  a user that owns $h$ blocks,
\begin{equation} 
G_h(n)=\{ (\bncustomers',0) + h \bon_n : \bncustomers' \in G(n-1) \}    \label{ghnn1}
\end{equation}
$\bon_n$ denotes a $1 \times n$ vector in which all elements equal zero, except element $n$, which equals one.  Let $p(\realav|{\bncustomers'})$ be the probability of $\realav$ blocks being \sustained{} when the upper layer state is $\bncustomers'$.  If $\bncustomers' \in G_h(n)$ ($n > 0$),   

\begin{equation} 
p(\realav|{\bncustomers'})=\sum_{k=0}^{B} p(k|{\bncustomers'-h \bon_n})     \psi_h(k,\realav), \qquad \bncustomers' \in G_h(n) \label{cbvm}
\end{equation}

According to definition  ~\eqref{umpvm}, 

\begin{equation}
\punsust_n(\realav)  {=}  \left[ \sum_{\bncustomers' \in G(n)} p(\realav|{\bncustomers'})   \pi(\bncustomers')\right] \Big/ \left[ \sum_{\bncustomers' \in G(n)} \pi(\bncustomers')  \right] = \sum_{\bncustomers' \in G(n)} p(\realav|{\bncustomers'})       \frac{1      }{ B^n \prod_{l=0}^{\nchunks-1}  {\ncustomers_l(\bncustomers')!}}
\end{equation}
where the first and second equalities follow from \eqref{unionb} and \eqref{condprobv}, respectively. 
Hence, 

\begin{equation}
p_n(\realav)=  \sum_{h=0}^{B-1} \sum_{\bncustomers'\in G_h(n)} p(\realav|{\bncustomers'})       \frac{1      }{ B^n \prod_{l=0}^{\nchunks-1}  {\ncustomers_l(\bncustomers')!}} {=}  \sum_{h=0}^{B-1} \sum_{\bncustomers' \in G_h(n)} \sum_{k=0}^{B} p(k|{\bncustomers'-h \bon_n})     \psi_h(k,\realav)   \frac{1      }{ B^n \prod_{l=0}^{\nchunks-1}  {\ncustomers_l(\bncustomers')!}}
\end{equation}

\begin{equation} \label{lasteq}
{=} \sum_{h=0}^{B-1}  \sum_{k=0}^{B}  \underbrace{ \sum_{\bmm'\in G(n-1)} p(k|{\bmm'})      \frac{1}{ B^{n-1} \prod_{l=0}^{\nchunks}  {\ncustomers'_l!}} }_{\punsust_{n-1}(k)} \psi_h(k,\realav) \frac{1}{B}   
{=} \sum_{h=0}^{B-1}  \sum_{k=0}^{B}  \punsust_{n-1}(k) \psi_h(k,\realav) \frac{1}{B}
\end{equation}
where the first, second and third equalities follow from \eqref{unionb}, \eqref{cbvm} and {\eqref{ghnn1}}, respectively.   Note that the summands in  ~\eqref{lasteq} for which  $h > v$  or $k \in [0, v-h-1] \cup [v+1, B]$ are equal to zero, since in these cases $\psi_h(k,v) = 0$. \eat{ ${m \choose n}$ =0 if $m > n$.  }  Therefore,  ~\eqref{lasteq} yields~\eqref{mainrec}.  \eat{Note that the summation limits in ~\eqref{mainrec} differ from those in  limits  ignoring the terms  that are equal to zero.   }

\section{Probabilistic Derivation of ~\eqref{1okp1}}

\label{sec:appderiv1ok}

Next, we provide a probabilistic derivation of~\eqref{1okp1}. An algebraic proof can be obtained using the Sigma package~\cite{sigma}, applying a paradigm called creative telescoping~\cite{doron}.

Henceforth, we consider the case $\realav < \nchunks$ (the case $\realav=\nchunks$ follows similarly).   The probabilistic interpretation for~\eqref{1okp1} follows from the connection between $\psi_h(k,\realav)$ and the  hypergeometric distribution.  Suppose we have an urn containing $B$ balls, $B-k$ of which are white (unavailable blocks) and $k$ are black (available blocks).   $\psi_h(k,\realav)$ is the probability of selecting without replacement $h$ balls, of which $\realav-k \leq h$ are white.

Consider now another experiment, namely selecting without replacement all balls from the urn. Let $J_w$ be the round in which the  $w^{th}$ white ball is selected, $0 < J_1 < J_2 < \ldots < J_{B-k} <  \nchunks+1$.
Let $W_b$ be the number of black balls selected between the $b^{th}$ and $b+1^{th}$ white, plus one.  
Equivalently, $W_b$ is the number of elements in the set 
$S=\{ n \in \{ 0, \ldots, \nchunks \} :\textrm{ exactly $b$ white balls are selected before ball $n$} \}$.   Then $W_0=J_1$, $W_1=J_2-J_1$, $\ldots$, $W_{k-1}=J_k-J_{k-1}$, and $W_{B-k}=\nchunks+1-J_i$. Clearly, 
$
\sum_{b=0}^{B-k} W_b = \nchunks+1
$. 
  By symmetry, $E[W_b] = (\nchunks+1)/(B-k+1)$ ($0 \leq b \leq  B-k$).   

Now let $\realav-k$ be given,  $\realav-k \in \{ 0, 1, \ldots, \realav \}$.  Let $\bon_h$ be the indicator equal to 1 if exactly $\realav-k$ white balls are selected among the first $h$ balls.  Note that $E[\bon_h] = \psi_h(k,\realav)$. Also, from the definition of $W_{\realav-k}$, we have $W_{\realav-k} = \sum_{h=0}^{\nchunks} \bon_h$.  
Therefore, $E[W_{\realav-k}] = \sum_{h=0}^{\nchunks} E[\bon_h] =\sum_{h=0}^{\nchunks}  \psi_h(k,\realav)  = (\nchunks+1)/(B-k+1)$.

\section{Proof of Theorem~\ref{theo:efficientalgo}}

\label{sec:appeff}

Replacing~\eqref{1okp1} into ~\eqref{simplified1} yields,

\begin{equation} 
\punsust_n(\realav) {=} \left\{  
\begin{array}{lll}  
1/B^n, & n \ge 1, \realav=0 &  (i) \\
\sum_{k=0}^{\realav} p_{n-1}(k)  (B+1)/(B(B-k+1)), & n \ge 1,  0 < \realav  < B & (ii) \\
 1-\sum_{l=0}^{B-1}  \punsust_n(l), &   n \ge 1, \realav = B  & (iii)  \\
1, & n =0, \realav = 0 &  (iv)  \\
0, & n =0, \realav \neq 0 &  (v)  
\end{array} \right.
\end{equation}
In case  $(ii)$, $p_n(\realav)=p_{n-1}(\realav) (B+1)/(B(B-\realav+1))  + \sum_{k=0}^{\realav-1} p_{n-1}(k)  (B+1)/(B(B-k+1)) = p_{n-1}(\realav) (B+1)/(B(B-\realav+1))  +  p_{n}(\realav-1)$ which yields~\eqref{dimsimple}.

\section{Expression of $p_n(\realav)$  When $\gamma=\infty$}

\label{sec:apppnv}

If $\gamma=\infty$, the closed-form expression for $p_n(\realav)$ is
\eat{\begin{equation} \label{closedformextension}
\punsust_n(\realav) {=} \left\{  
\begin{array}{ll}  
\eat{ (1/B)^n, & n \ge 1, \realav=B  \\ }
{B \choose \realav} \left(\frac{1+B}{B}\right)^n \sum_{l=0}^{\realav} {\realav \choose l} (-1)^l (B-\realav+l+1)^{-n}, & n \ge 1, B > \realav \ge 0 \\
1 - \sum_{\realav=0}^{B-1} \punsust_n(\realav), & \realav = B
\end{array} \right.
\end{equation}
}
$\punsust_n(\realav) {=} {B \choose \realav} \left(\frac{1+B}{B}\right)^n \sum_{l=0}^{\realav} {\realav \choose l} (-1)^l (B-\realav+l+1)^{-n}$,  $n \ge 1, B > \realav \ge 0$, 
 and $\punsust_n(B) = 1 - \sum_{\realav=0}^{B-1} \punsust_n(\realav)$.

\section{Self-Sustainability with Heterogeneous Download Times}

\label{sec:apphet}

We now present an efficient algorithm to compute swarm self-sustainability if the block download times, $\mu_h$, are not necessarily equal for all $h$, $1 \leq h \leq B$.  

Denote by $a_{h,r}(\realav)$
the probability that $\realav$ blocks are \sustained{} conditioned on the presence of $r$ users in layer $h$ and all users being in layers $h$ up to $B$,

\begin{equation} \label{umpvm2}
a_{h,r}(\realav) = P(V=\realav \Big| \ncustomers_h=r; \ncustomers_i = 0, i < h ), \qquad 0 \leq h \leq B, \qquad 0 \leq r \leq M
\end{equation}
$M$ is the maximum number of users per layer.  
The  following recursion correctly computes $a_{h,r}$, $0 \leq h \leq B$, $0 \leq r \leq M$,
\begin{equation} \label{ahr}
a_{h,r}(\realav){=}\left\{ \begin{array}{lll}
  {\displaystyle{\sum_{\substack{i=0}}^{\realav}}}  a_{h,r{-}1}( i) \psi_h(i,\realav), & {  r \ge 1 }, h\neq B & (i) \\
 \sum_{r=0}^{\infty} a_{h+1,r}(\realav)\pi_{h+1}(r), & r=0, h\neq B & (ii) \\
 1, & r=0, h=B, v=0  & (iii)\\
 0, & r=0, h = B, v\neq0 & (iv)\\
 \end{array} \right.
\end{equation}
We approximate $A$ \eat{the self-sustainability  } by its truncated version, $\overline{A}^{(M)}$, considering only  states in which there are up to $M$ users in each layer of the system.      Recall that $N$ is the maximum number of users in the system, $N=O(BM)$. A naive  use of~\eqref{ahr} yields an algorithm to compute $\overline{A}^{(M)}$ in time $O(B^3 M)=O(B^2 N)$.

In what follows, we show how to compute $\overline{A}^{(M)}$ in $O(N B \log B)$. For this purpose, we re-write the sum in ~\eqref{ahr}$(i)$ as a convolution, which is computed in time $O(B \log B)$.
Let
$$\widehat{a}_{h,r{-}1}(k) =  a_{h,r{-}1}(k) {k!} {(B{-}k)!}, \quad 0 \leq k \leq B \quad \textrm { and } \quad \widehat{\psi}_h(\realav-k)=1/({(\realav{-}k)!}  {(h{-}(\realav{-}k))!}), \quad  0 \leq k \leq \realav $$

Then, the following convolution correctly computes ~\eqref{ahr}$(i)$,
$
\widehat{{{{a}}}}_{h,r} =   \widehat{{{{a}}}}_{h,r-1} \otimes \widehat{{\psi}}_h $.

 ${a}_{h,r}$ is obtained from $\widehat{a}_{h,r}$ as
\begin{equation}
a_{h,r}(\realav) = 
\left\{ \begin{array}{ll} ({\widehat{a}_{h,r}(\realav)})/\left({  (\realav-h)! (B{-}\realav)!{{\nchunks \choose h}}}\right), &  0 \leq \realav \leq B-1 \\
1-\sum_{i=0}^{\nchunks-1} a_{h,r}(i), & v = B\\
\end{array} \right.
\end{equation}

\label{sec:apprecalt}

\section{Derivation of ~\eqref{closedform}}
\label{sec:appderclosed}

We now show that~\eqref{closedform} follows from~\eqref{dimsimple2}. 
The case $\realav=B$ follows trivially.  For $\realav  < B$, we show that~\eqref{closedform}
satisfies~\eqref{dimsimple2}, i.e., $
\punsust_n(\realav)- \punsust_{n-1}(\realav)(1/(B-\realav+1)) =\punsust_n(\realav-1)$,
\begin{equation} \nonumber
\punsust_n(\realav)- \punsust_{n-1}(\realav)(1/(B-\realav+1))  = {B \choose \realav} \sum_{l=0}^{\realav} {\realav \choose l} (-1)^l (B-\realav+l+1)^{-n} - \frac{ {B \choose \realav} \sum_{l=0}^{\realav} {\realav \choose l} (-1)^l (B-\realav+l+1)^{-n+1} }{B-\realav+1}
\end{equation}
\begin{equation} \nonumber
= {B \choose \realav} \sum_{l=0}^{\realav} {\realav \choose l} (-1)^{l+1} (B-\realav+l+1)^{-n}  \frac{l}{B-\realav+1} 
=  {B \choose \realav-1} \sum_{l=0}^{\realav-1} {\realav-1 \choose l} (-1)^{l} (B-\realav+l+2)^{-n}   = \punsust_n(\realav-1)
\end{equation}

\section{Derivation of~\eqref{d0}}

\label{sec:d0}

We now derive ~\eqref{d0} from $\punsust_n(B)= \sum_{l=0}^{B} {B \choose l} (-1)^l (l+1)^{-n} $.
\begin{eqnarray}
\punsust(B)&=&\sum_{n=0}^\infty \sum_{l=0}^{B} \punsust_n(B) e^{-(B+1) \rho} ((B+1) \rho)^n / n! =  \\
&=& \sum_{l=0}^{B}  {B \choose l} (-1)^l e^{-B \rho} e^{(B+1) \rho/(l+1)} \underbrace{\sum_{n=0}^\infty e^{-(B+1) \rho/(l+1)}   ((B+1) \rho/(l+1))^n / n!}_{=1} \nonumber
\end{eqnarray}

\section{Probability that $l$ tagged blocks are unavailable}

\label{taggedblocks}

The probability that $l$ tagged blocks are unavailable, when $\gamma=\mu$, is 
$\exp\left( -\rho \left(   \frac{ l(B+1) }{l+1 } \right) \right)$.

Let $\alpha_{h,r}$ be the probability that there are $l$ tagged unavailable blocks, conditioned on all users being in stages $h, h+1, \ldots, \nchunks-1, \nchunks$, stage $h$ having $r$ users (similar in spirit to recursion presented in Appendix~\ref{sec:apprecalt}).    $\alpha_{h}$ is the corresponding metric, but with no conditioning on the number of users in layer $h$.
Given $l$, the following recursion yields~$\alpha_{h,r}$.   
\begin{equation}  \label{eq:probltarecursion}
\alpha_{h,r}{=}\left\{ \begin{array}{rlr}
 (i) & 1 & h {=} \nchunks, r {=} 0 \\
(ii) &  \alpha_{h,r-1} \prod_{i=0}^{h-1} \frac{B-l-i}{B-i} & {  h \leq \nchunks, r \ge 1 }\\
(iii) & \sum_{m=0}^\infty  \alpha_{h+1,m} \pi_{h+1}(m) & h \leq \nchunks,  r=0 \\
 \end{array} \right. \end{equation}

Our goal is to find the expression of $\alpha_{h,0}$, 
 \begin{equation} \label{aux0}
 \alpha_{h,0} = \sum_{m=0}^{\infty} \alpha_{h+1,m} \pi_{h+1}(m)
 \end{equation}

From ~\eqref{eq:probltarecursion}$(ii)$, 
 \begin{equation} \label{eq:recah1m}
 \alpha_{h+1,m} =  \alpha_{h+1,m-1} \prod_{i=0}^{h} \frac{B-l-i}{B-i} \therefore  \alpha_{h+1,m} =  \alpha_{h+1,0} \Big( \prod_{i=0}^{h}  \frac{B-l-i}{B-i} \Big)^{m}
 \end{equation}
 Hence, replacing~\eqref{eq:recah1m} into \eqref{aux0}, 
  \begin{equation}
\alpha_{h,0} = \sum_{m=0}^{\infty} \alpha_{h+1,0} \Big(\prod_{i=0}^{h} \frac{B-l-i}{B-i}\Big)^m \frac{e^{-\rho} \rho^m}{m!}
  = \alpha_{h+1,0} \sum_{m=0}^{\infty}  \Big[ \underbrace{\prod_{i=0}^{h} \frac{B-l-i}{B-i}}_{\beta_{h+1}} \Big]^m \frac{e^{-\rho} \rho^m}{m!}
 \end{equation} 
 Let \eat{ 
 \begin{equation}
 \beta_{h+1} = \prod_{i=0}^{h} \frac{B-l-i}{B-i}= \frac{{B-h-1 \choose l}}{{B \choose l}} 
 \end{equation} }
  \begin{equation}
 \beta_{h} = \prod_{i=0}^{h-1} \frac{B-l-i}{B-i} =\frac{{B-h \choose l}}{{B \choose l}}
 \end{equation}
Therefore,
   \begin{equation}
\alpha_{h,0} = \alpha_{h+1,0} \sum_{m=0}^{\infty}  ( {\beta_{h+1}} \rho )^m \frac{e^{-\rho} }{m!} 
 = \alpha_{h+1,0} e^{-x} \sum_{m=0}^{\infty}  ( {\beta_{h+1}} \rho )^m \frac{e^{-\rho+x} }{m!} = \alpha_{h+1,0}  e^{-x}
 \end{equation} 
 where $-\rho+x=-\beta_{h+1} \rho$,  
 \begin{equation}
 x=\rho -\beta_{h+1} \rho = \rho(1-\beta_{h+1})
 \end{equation} 
So,
 \begin{equation}
 \alpha_{h,0} = \alpha_{h+1,0} \exp(-\rho (1-\beta_{h+1}))
\eat{ \end{equation} 
 \begin{equation}
 \alpha_{h,0} = \alpha_{h+1,0} \exp(-\rho (1-\prod_{i=0}^{h} \frac{B-l-i}{B-i}))
 \end{equation}
 \begin{equation}
 \alpha_{h,0} = \alpha_{h+1,0} \exp(-\rho (1-\frac{(B-l)!l! (B-h-1)!}{(B-l-h-1)!B!l!}))
\end{equation}
 \begin{equation} 
 \alpha_{h,0}} = \alpha_{h+1,0} \exp(-\rho (1-\frac{{B-h-1\choose l}}{{B \choose l}})) \label{eqaux10}
\end{equation}
\eat{

Remark:

In particular, if we substitute $h=B-1$ in the equation above we obtain
 \begin{equation}
 \alpha_{B-1,0} = \alpha_{B,0} \exp(-\rho (1-0)) = e^{-\rho}
\end{equation}
as desired.

Also, $\beta_B = 0$.

End of remark.
}

The solution of recursion ~\eqref{eqaux10} is

\begin{equation} 
\alpha_{h,0}{=}   \left\{ \begin{array}{lll}
 (i) & 1,  & h {=} \nchunks \\
(ii) &   \prod_{a=h}^{\nchunks-1} \exp( -\rho ( 1- \frac{{B-a-1\choose l}}{{B \choose l}} )), & 0 \leq h < B \\
 \end{array} \right. \end{equation}
If $h=0$,

 \begin{equation} \label{almostfinal}
 \alpha_{0,0} = \prod_{a=0}^{\nchunks-1} \exp( -\rho ( 1-  \frac{{B-a-1\choose l}}{{B \choose l}} ))
=  \exp( -\rho (  \left( \sum_{a=0}^{\nchunks-1} 1 \right) - \left(  \sum_{a=0}^{\nchunks-1}  \frac{{B-a-1\choose l}}{{B \choose l}} \right) ))
\end{equation}

The hockey stick pattern of the Pascal triangle, \eat{\url{http://ptri1.tripod.com/}}
$\sum_{a=0}^{\nchunks-1}  {{B-a-1\choose l}} = {{B\choose l+1}}$, together with ~\eqref{almostfinal}, yields
 \begin{equation}
 \alpha_{0,0} = \exp\left( -\rho \left( B-  \frac{{B\choose l+1}}{{B \choose l}} \right)\right)
\eat{
 \begin{equation}
 \alpha_{0,0} = \exp\left( -\rho \left( B-  \frac{B! (B-l)! l!}{(B-l-1)! (l+1)! B!} \right) \right)
\end{equation}

 \begin{equation}
 \alpha_{0,0} = \exp\left( -\rho \left( B-  \frac{ B-l }{l+1 } \right) \right)
\end{equation}

 \begin{equation}
 \alpha_{0,0} = \exp\left( -\rho \left(   \frac{ B(l+1)-B+l }{l+1 } \right) \right)
\end{equation}
}
 = \exp\left( -\rho \left(   \frac{ l(B+1) }{l+1 } \right) \right)
\end{equation}

\section{Integrating the User Dynamics into the Lower Layer}

\label{app:integr}

Next, we provide an alternative  description of our model which explicitly characterizes the evolution of the signature of each user in time.  

Assume that each user selects uniformly at random one of the blocks among those that he does not have.  Each user then contacts other users uniformly at random for opportunities to download the selected block.  The time between contacts initiated by a specific (tagged) user is  characterized by a Poisson process of rate $\mu$.  If the contacted user has the requested block, it is transfered immediately. Otherwise, the tagged user is instantaneously re-directed to the publisher, who then transfers the block.   All transfers are assumed to be instantaneous.    

The system described above can be fully characterized  using the signatures introduced in our lower layer model.  Let $S$ be the lower layer state, i.e., $S$ is a $\{0,1\}^{Bn}$  bit vector in case there are $n$ users in the system, (see~\textsection\ref{sec:lowlayer}).   $n_i(S)$ is the number of users in stage $i$ when the system state is $S$.  Let $h(S,u)$ be the stage of user $u$ when the system state is $S$.

Let   $\bOmega_n$ be the union of the state spaces of the  lower layer model corresponding to an the upper layer in which there are $n$ users in the system, $\bOmega_n = \bigcup_{\bncustomers:|\bncustomers|=n} \bOmega_{|\bncustomers|}$.     Let $\bOmega'$ be the union of all possible state spaces of the lower layer model, $\bOmega'= \bigcup_{i=0}^{\infty} \bOmega_i$.  
Starting from state $S$, let $T_0$ be the state resulting from an arrival and  $T_{u,b}$  be the state resulting  from user $u$  concluding the download of  block $b$.

Let $\bzr$ denote a vector of lenght $B$ with all elements equal to zero;  $\be_i$ denotes a vector with its $i^{th}$ element equal to 1 and all other elements equal to zero;  $S \setminus u$ denotes the bit vector $S$ after the removal of the bits corresponding to user $i$ (bits $(u-1)B+1$ up to $uB$).   We assume that after a user leaves the system, the remaining users are re-indexed accordingly. 

\begin{eqnarray}
T_0: \bOmega_n &\rightarrow& \bOmega_{n+1} \nonumber \\
S &\rightarrow& S.\bzr \\ 
\end{eqnarray}

If $h(S,u) < B-1  $
\begin{eqnarray}
T_{u,b}: \bOmega_n &\rightarrow& \bOmega_{n} \nonumber\\ 
S &\rightarrow& S + \be_{B(u-1)+b}  \label{eq:case1}
\end{eqnarray}

If $h(S,u) = B-1  $
\begin{eqnarray}
T_{u,b}: \bOmega_n &\rightarrow& \bOmega_{n-1} \nonumber\\ 
S &\rightarrow&  S \setminus u \label{eq:case2}
\end{eqnarray}

Note that transformation $T_{u,b}$ explicitly characterizes the evolution of the signature of each user in time.    \eqref{eq:case1} corresponds to the download of a block  and a signature update whereas \eqref{eq:case2} corresponds to a download completion.

The entries of the generator matrix $Q=(q(S,S'): S,S' \in \bOmega')$ are 
\begin{eqnarray}
q(S,T_0(S)) &=&  \lambda \label{eqfinal1} \\ 
q(S,T_{u,b}(S)) &=&  \mu/(B-h(S,u)) \label{eqfinal2} 
\end{eqnarray}
Equations  \eqref{eqfinal1}-\eqref{eqfinal2} provide an integrated description of the upper and lower layer models.  This model description is similar in spirit to~\cite{syndrome}.  While Hajek and Zhu~\cite{syndrome} are  interested in studying the \emph{stability} of a system that resembles the one described above, the system considered in this paper is always stable, and we are interested in its \emph{self-sustainability}.

\end{appendix}







\end{document}
